\documentclass[%
 reprint,
nofootinbib,
floatfix,
superscriptaddress]{revtex4-1}
\usepackage[english]{babel}
\usepackage{siunitx}
\usepackage{dcolumn}

\usepackage[caption=false]{subfig}
\usepackage{floatrow}
\usepackage{amsfonts}
\usepackage{microtype}
\usepackage{amsmath}
\usepackage{mathtools}
\usepackage{tabularx}
\usepackage{appendix}
\usepackage{chngcntr}
\usepackage{graphicx}
\usepackage{dcolumn}
\usepackage{bm}
\usepackage[colorlinks=true, allcolors=blue, pdfencoding=auto]{hyperref}

\usepackage[Symbolsmallscale]{upgreek}

\DeclareMathAlphabet\mathbfcal{OMS}{cmsy}{b}{n}

\addto\extrasenglish{%
}

\addto\extrasenglish{%
}

\addto\extrasenglish{%
}

\addto\extrasenglish{%
}

\addto\extrasenglish{%
}
\addto\extrasenglish{%
}

\usepackage[capitalize,nameinlink]{cleveref}
\usepackage[integrals]{wasysym}
\creflabelformat{equation}{#2#1#3}
\crefname{section}{Sec.}{Secs.}
\DeclareMathAlphabet{\mathpzc}{OT1}{pzc}{m}{it}

\begin{document}


\title{Atmospheric Lepton Fluxes via Two-Dimensional Matrix Cascade Equations}

\author{Tetiana Kozynets}%
 \email{tetiana.kozynets@nbi.ku.dk}
\affiliation{Niels Bohr Institute, University of Copenhagen, Blegdamsvej 17, 2100 Copenhagen, Denmark}
\author{Anatoli Fedynitch}
\affiliation{Institute of Physics, Academia Sinica, Taipei City, 11529, Taiwan}
\affiliation{Institute for Cosmic Ray Research, University of Tokyo, 5-1-5 Kashiwa-no-ha, Kashiwa, Chiba 277-8582, Japan}
\author{D. Jason Koskinen}
\affiliation{Niels Bohr Institute, University of Copenhagen, Blegdamsvej 17, 2100 Copenhagen, Denmark}

\date{\today}

\begin{abstract}
The atmospheric lepton fluxes play a crucial role in many particle and astroparticle physics experiments, e.g.\@ in establishing the neutrino signal and the muon background for neutrino oscillation measurements, or the atmospheric background for astrophysical neutrino searches. The Matrix Cascade Equations (\textsc{MCEq}) code is a numerical tool used to model the atmospheric lepton fluxes by solving a system of coupled differential equations for particle production, interaction, and decay at extremely low computational costs. Previously, the \textsc{MCEq} framework accommodated only longitudinal development of air showers, an approximation that works well for neutrino and muon fluxes at high energies ($\mathcal{O}(\SI{10}{GeV})$ and above). However, for accurate calculations of atmospheric lepton angular distributions at lower energies, the lateral component of hadronic cascades becomes significant. We introduce ``2D \textsc{MCEq}'', an efficient numerical approach for combined longitudinal and angular evolution of air showers that retains the low computational complexity. The accuracy of the 2D \textsc{MCEq} is affirmed by its benchmark comparison with the standard Monte Carlo code \textsc{corsika}. Our method can be used for two-dimensional evolution of hadronic cascades in arbitrary media and paves the way for efficient three-dimensional calculations of atmospheric neutrino fluxes.

\end{abstract}

\maketitle

\section{Introduction}\label{sec:intro}

Interactions of cosmic rays with the atomic nuclei in the Earth's atmosphere produce cascades of secondary particles, referred to as the extensive air showers. These cascades have two components -- electromagnetic (encompassing production and subsequent reinteraction of energetic electrons and photons) and hadronic (including production and subsequent reinteraction/decay of unstable mesons). One of the byproducts of hadronic cascades are neutrinos, which span the entire energy range from MeV to PeV and thereby form a broad landscape for probing fundamental physics. In particular, the GeV-scale neutrinos produced in the air showers constitute the main signal for atmospheric neutrino oscillation studies, including e.g.\@ muon neutrino disappearance \cite{IceCube:2014flw, IceCube:2017lak, Super-Kamiokande:2017yvm, MINOS:2014rjg}, tau neutrino appearance \cite{IceCube:2019dqi, Super-Kamiokande:2017edb, Super-Kamiokande:2006yyp, Eberl:2017plv, Hallmann:2021jbk}, and searches for physics beyond the Standard Model \cite{Stuttard:2020qfv, Coloma:2018idr, Schneider:2021wzs, Fischer:2022zwu}. In addition, atmospheric neutrinos are an ``irreducible background'' for astrophysical neutrino searches (e.g.\@ \cite{IceCube:2013low, IceCube:2016umi, IceCube:2020qls, IceCube:2022lnv}), which further strengthens the motivation for accurate modelling of neutrino production in the Earth's atmosphere. 

The unoscillated neutrino fluxes depend on several main inputs: the primary cosmic ray flux (including composition and spectrum \cite{Workman:2022ynf,Gaisser:2012sdr,Dembinski:2017zsh, Gaisser:2013bla}); the hadronic interaction model (prescribing the probabilities of secondary particle yields in hadron-nucleus collisions; e.g.\@ \cite{Fedynitch:2018cbl, Riehn:2019jet, Pierog:2013ria, Bass:1998ca, Roesler:2000he,Fedynitch:2015kcn}); decay probabilities and branching ratios of unstable particles \cite{Bierlich:2022pfr, Workman:2022ynf}; and model for the atmospheric density as a function of altitude for specific geographical locations \cite{Picone:2002nrl, Gaisser:2016crp, Honda:2015fha, Heck:1998vt}. At $\mathcal{O}$(GeV) and sub-GeV energies, the angular distributions of the air shower secondary particles (``secondaries'') are further affected by the Earth's magnetic field, which curves the trajectories of the charged cosmic ray primary particles (``primaries'') and the secondary muons. In addition, the angular spread of the low-energy secondaries with respect to the primary particle axis becomes non-negligible: as the transverse momentum is Lorentz-invariant, the deflection angle grows with decreasing energy and can vary from a few degrees to tens of degrees at GeV-scale energies. Both of these effects are necessary to properly describe the angular evolution of individual air showers and the resulting full-sky angular distribution of the $\mathcal{O}$(GeV) atmospheric neutrinos \cite{Lipari:2000wu}.

Monte Carlo simulations are the most natural approach to incorporate the many stages of the air shower modelling into a single computational framework \cite{Honda:2004yz, Barr:2004br}. The Monte Carlo treatment implies that the generation and propagation of the cosmic ray primaries and the interactions and decays of the secondaries are executed on an event-by-event basis. These processes are stochastic and follow the probabilistic particle yield prescriptions of a given \textit{event generator}. The most widely used realizations of this method to date include the general-purpose codes such as \textsc{Geant4} \cite{GEANT4:2002zbu}, \textsc{fluka} \cite{Battistoni:2007zzb}, \textsc{mcnp} \cite{osti_1419730} and \textsc{phits} \cite{PHITS:2018} for particle propagation in matter, as well as \textsc{corsika} \cite{Heck:1998vt} and \textsc{aires} \cite{Sciutto:1999jh} codes specialized in air shower evolution. The closed-source HKKMS \cite{Honda:2004yz} and Bartol \cite{Barr:2004br} atmospheric neutrino flux models are also based on Monte Carlo simulations, employing event generators \textsc{Jam}+\textsc{DPMJet} \cite{Honda:2006qj,Honda:2011nf, Nara:2019epj,Roesler:2000he} and \textsc{Target} \cite{Engel:2003bv}, respectively. The HKKMS model has been tuned to reproduce the muon flux data \cite{Honda:2006qj,Honda:2019ymh} and is set out to incorporate the fixed-target experiment measurements of hadronic interaction yields in the future \cite{Sato:2021vfq}. It is therefore commonly used as the baseline atmospheric neutrino flux model in experimental analyses by e.g.\@ the Super-Kamiokande and the IceCube collaborations, as well as in projections for the upcoming Hyper-Kamiokande, JUNO, and DUNE experiments \cite{JUNO:2021tll, Fukasawa:2016yue}. Both the HKKMS and the Bartol flux models include the geomagnetic effects and the deflection of the secondaries from the primary axis, and therefore are the standard reference for $\mathcal{O}$(GeV) atmospheric lepton flux calculations. While the Monte Carlo approach for the inclusive atmospheric lepton flux calculations provides high level of detail as an inherent advantage, it is computationally expensive, fairly complex, and lacks sufficient flexibility for extraction of systematic uncertainties (e.g.\@ those related to the cosmic ray flux model and the hadronic interaction model parameters \cite{Barr:2006it, Fedynitch:2022vty, Dembinski:2017zsh}).

Another natural path towards the inclusive atmospheric neutrino flux modelling is via a solution to the cascade equations describing particle production, interaction, and decay in the atmosphere (see e.g.\@ \cite{Gaisser:2016crp} for a broad introduction into this topic). Numerous studies have tackled these equations semi-analytically, with \cite{Naumov:2000au, Kochanov:2008pt, Lipari:1993hd} being the latest developments. The semi-analytical method was further overtaken by the high-precision numerical solutions provided by the  \textsc{MCEq} software \cite{Fedynitch:2015zma, Fedynitch:2018cbl}. The \textsc{MCEq} cascade equation solver relies on the probabilities of the secondary particle yields in the interaction and decay processes extracted from event generators and stored as matrices. Avoiding the need to repeatedly run event generators within the user interface, \textsc{MCEq} enables computation of inclusive secondary particle fluxes on millisecond timescales, compared to several CPU-hours typically required by the Monte Carlo calculations. 

Despite the significant speedup over the Monte Carlo approaches and the flexibility to study the impact of the systematic parameters, the \textsc{MCEq} code could not be readily used to predict the angular distributions of the $\mathcal{O}$(GeV) cascade secondaries. The reason for this constraint is that \textsc{MCEq} was originally written in the 1D approximation of the air shower development, i.e., under the assumption of strictly collinear (with respect to the primary cosmic ray axis) secondary particle production and propagation. 

In this study, we are seeking to extend the \textsc{MCEq} framework with the angular evolution of the individual air showers. We develop the numerical technique and the practical implementation of a two-dimensional cascade equation solver, where the secondaries are allowed to deviate from the primary particle trajectory. Our code, ``2D \textsc{MCEq}''\footnote{\href{https://github.com/kotania/MCEq/tree/2DShow/}{https://github.com/kotania/MCEq/tree/2DShow}}, enables numerical computation of the resulting angular distributions of secondary particles. This advancement has broad applications in the analyses involving $\mathcal{O}$(GeV) atmospheric leptons (and more generally, any hadronic cascade secondaries in arbitrary media) and contributes to future development of fully numerical or hybrid three-dimensional calculations of atmospheric neutrino fluxes and air showers.

This paper is structured as follows. In \autoref{sec:1d_equations}, we review the analytical cascade equations in the one-dimensional approximation and further show how to incorporate the second (angular) dimension in \cref{sec:extending_to_2d,sec:analytical_twodim_equations}. The numerical (matrix) form of the 1D equations forming the basis of the \textsc{MCEq} code is reviewed in \autoref{sec:1d_equations_matrix}. We then derive the matrix form for the 2D equations by reformulating them in the frequency domain (\autoref{sec:hankel_space_equations}). The pipeline of the 2D \textsc{MCEq} software is described in \autoref{sec:2d_mceq_pipeline}, which includes the steps to prepare the interaction/decay probability matrices (\cref{sec:event_generation,sec:2d_matrices}) and the principles of the 2D cascade equation integrator (\cref{sec:hankel_solution,sec:solution_reconstruction}). Finally, we compare the 2D \textsc{MCEq} angular distributions to those obtained with the \textsc{corsika} Monte Carlo in \autoref{sec:benchmarking}, focusing on 2D cascades induced by a single cosmic ray primary.

\section{Cascade equation theory}\label{sec:cascade_eq}

\subsection{One-dimensional cascade equations}\label{sec:1d_equations}

In this work, we employ the cascade theory to characterize the spatial development of the secondary particle showers induced by a single cosmic ray projectile. The mathematical basis of this theory is the system of coupled partial integro-differential \textit{cascade equations}, which are a form of the Boltzmann transport equations for multiple particle species. For the secondary particle species $h$, we define the single-differential particle density $n_h$ with respect to the kinetic energy $E$:
\begin{equation}
    n_h(E) = \frac{\mathrm{d}N_h}{\mathrm{d}E},
\end{equation}
which represents the number of particles $N_h$ per energy interval. In the one-dimensional cascade theory, this single-differential density evolves as a function of the atmospheric slant depth $X$: 
\begin{equation}
    X(h_{\mathrm{o}}) = \int_{0}^{h_{\mathrm{o}}} \mathrm{d}l \rho_{\mathrm{air}}(l),
\label{eq:slant_depth_definition}
\end{equation}
where $h_{\mathrm{o}}$ is the observation altitude above the surface of the Earth, $\rho_{\mathrm{air}}$ is the depth-dependent air density, and the integral is evaluated along the trajectory $l$ of the shower core. With $\rho_{\mathrm{air}}$ given in $\mathrm{g\,cm^{-3}}$, and $l$ taken in $\mathrm{cm}$, the unit of $X$ is $\mathrm{g\,cm^{-2}}$.
Then, the one-dimensional coupled cascade equations \cite{Gaisser:2016crp, Fedynitch:2018cbl, Fedynitch:2015zma} read 
\begin{subequations}
\begin{align}
    \frac{\mathrm{d} n_h(E, X)}{\mathrm{d}X} = & -\frac{n_h (E, X)}{\lambda_{\mathrm{int}, h}(E)} -\frac{n_h (E, X)}{\lambda_{\mathrm{dec}, h}(E, X)} \label{eq:ana_onedim_cascade_sinks}\\
    &-\frac{\partial}{\partial E} (\mu_E n_h(E, X))
    \label{eq:ana_onedim_energy_losses}\\
    & + \sum_{\ell} \int_{E}^{\infty}\mathrm{d}E_{\ell}\,\frac{\mathrm{d}N_{l(E_{\ell}) \to h(E)}^{\mathrm{dec}}}{\mathrm{d}E} \frac{n_{\ell} (E_{\ell}, X)}{\lambda_{\mathrm{int}, {\ell}}(E_{\ell})}  \label{eq:ana_onedim_cascade_inter_sources}\\
    & + \sum_{\ell} \int_{E}^{\infty}\mathrm{d}E_{\ell}\,\frac{\mathrm{d}N_{{\ell}(E_{\ell}) \to h(E)}^{\mathrm{int}}}{\mathrm{d}E} \frac{n_l (E_{\ell}, X)}{\lambda_{\mathrm{dec}, {\ell}}(E_{\ell}, X)} \label{eq:ana_onedim_cascade_decay_sources}. 
\end{align}
\label{eq:ana_onedim_cascade_eq}
\end{subequations}
The ``sink'' terms in \cref{eq:ana_onedim_cascade_sinks} represent the decrease in the density of particle type $h$ as the result of its interactions in the atmosphere after travelling the interaction length $\lambda_{\mathrm{int}, h}$, or its decay after travelling the decay length $\lambda_{\mathrm{dec}, h}$. Another sink term in \cref{eq:ana_onedim_energy_losses} stands for the energy losses of the charged particles due to ionization, where ${\mu_E = -\langle \frac{\mathrm{d}E}{\mathrm{d}X} \rangle}$ is the average stopping power per unit length. The ``source'' terms in \cref{eq:ana_onedim_cascade_inter_sources} and \cref{eq:ana_onedim_cascade_decay_sources} describe the increase of $n_h$ due to the interactions and decays of other particle species ${\ell}$ with energy $E_{\ell}$. The respective yields of the particle $h$ are reflected in the differential production cross sections $\frac{\mathrm{d}N_{{\ell}(E_{\ell}) \to h(E)}}{\mathrm{d}E}$. The energy conservation constraint is given in the integral bounds ($\int_{E}^{\infty}$) of \cref{eq:ana_onedim_cascade_inter_sources,eq:ana_onedim_cascade_decay_sources}: it requires that the total energy $E_{\ell}$ of the primary particle must be greater than, or equal to, the total energy $E$ of the secondary particle.

\subsection{Incorporating the second (angular) dimension}\label{sec:extending_to_2d}

In high-energy inelastic collisions or decays, the angular deflection $\theta_{\ell \to h}$ of the secondary particles $h$ from the primaries $\ell$ is minor ($\ll 1^{\circ}$ at energies $\gg$\SI{10}{GeV}), justifying the use of the 1D approximation in the evolution of high-energy hadronic cascades \cite{Lipari:2000wu}. In this regime, the velocity unit vector $\hat{\mathbf{u}}_{{\ell} \rightarrow h}$ of $h$ translates to $( 0, 0, 1 )^{\top}$ in a Cartesian coordinate system where the $z$ axis aligns with ${\ell}$. Lower energies necessitate consideration of the $x$ and $y$ components of $\hat{\mathbf{u}}_{{\ell} \rightarrow h}$ and explicit inclusion of the azimuthal angle $\varphi_{\ell \rightarrow h}$. Then, the velocity vector becomes $\hat{\mathbf{u}}_{{\ell} \rightarrow h} = (\sin \theta_{{\ell} \rightarrow h} \cos \varphi_{{\ell} \rightarrow h},\,\sin \theta_{{\ell} \rightarrow h} \sin \varphi_{{\ell} \rightarrow h}, \,\cos \theta_{{\ell} \rightarrow h})^{\top}$. To second order in $\theta$, this can be approximated as 
\begin{equation}
    \hat{\mathbf{u}}_{{\ell} \rightarrow h} = \begin{pmatrix} \theta_{{\ell} \rightarrow h} \cos \varphi_{{\ell} \rightarrow h} \\ \theta_{{\ell} \rightarrow h} \sin \varphi_{{\ell} \rightarrow h} \\  1 - \frac{(\theta_{{\ell} \rightarrow h})^2}{2} \end{pmatrix}
    \label{eq:unit_vector_relative}.
\end{equation} 
Similarly, the initial particle ${\ell}$ can be assigned a unit velocity vector $\hat{\mathbf{u}}_{{\ell}} = \left(\theta_{{\ell}} \cos \varphi_{\ell},\, \theta_{\ell} \sin \varphi_{\ell}, \,1 - \frac{(\theta_{\ell})^2}{2}\right)^{\top}$ in a fixed-frame Cartesian coordinate system, where $\theta_{\ell}$ is the angle between the direction of $\ell$ and the $z$ axis of this system, and $\varphi_{\ell}$ is the respective azimuthal angle.

\begin{figure*}[htb!]
  \includegraphics[width=\textwidth]{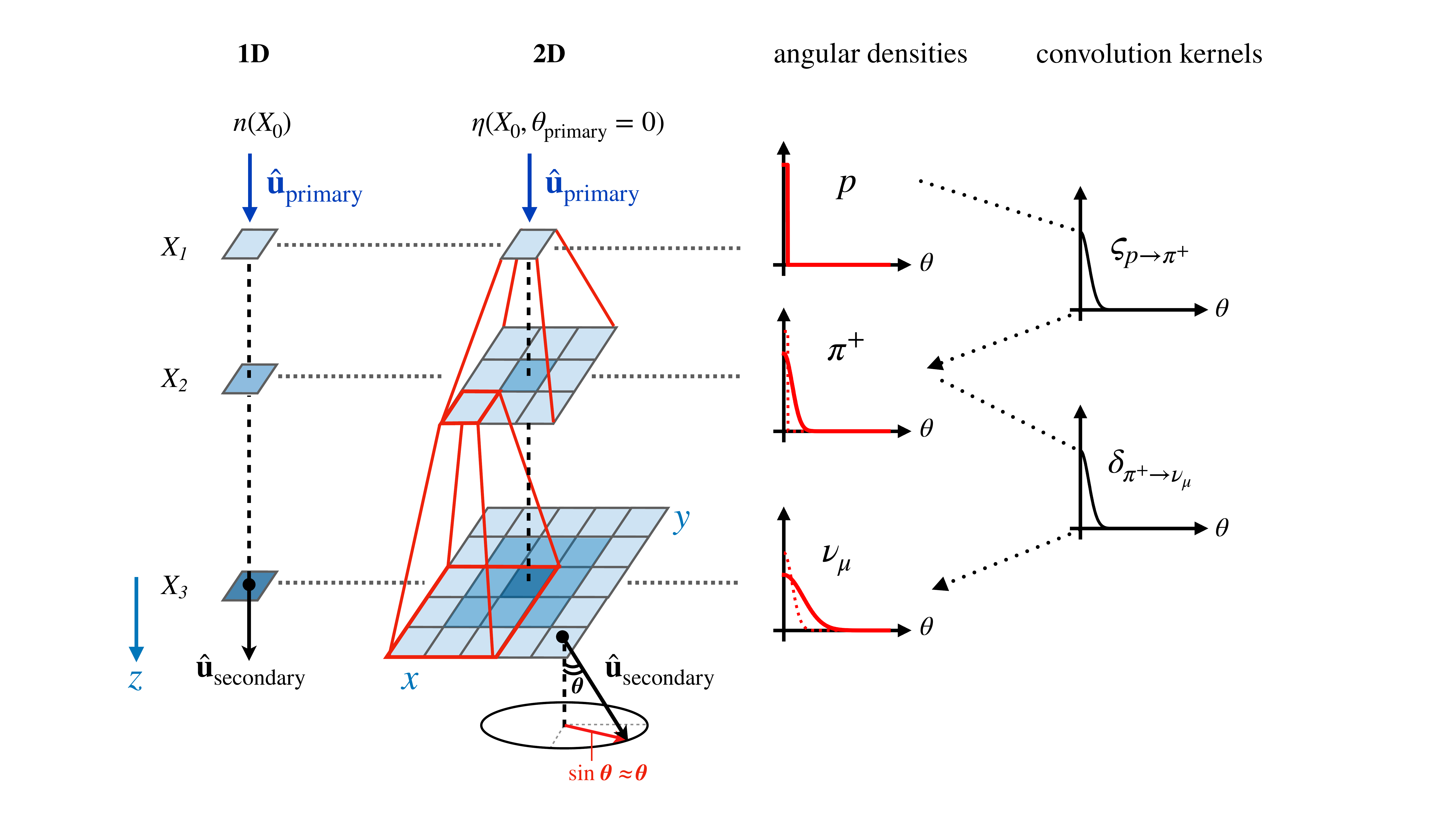}
  \caption{Schematic development of a hadronic cascade ($p$ $\to$ $\pi^{+}$ $\to$ $\nu_{\mu}$) in the 1D (longitudinal-only) and the 2D (longitudinal + angular) geometries. In this diagram, the longitudinal propagation is performed over three discrete steps along the slant depth $X$ for illustrative purposes. At each step in $X$, the angular distribution of the primaries from the previous step is shown as the dotted line, and the current angular distribution of the specified particle as the solid line. The distributions of secondaries get wider further down the chain due to the convolution with the kernels $\varsigma_{p \to \pi^{+}}$ and $\delta_{\pi^{+} \to \nu_{\mu}}$ (see text for details).}
  \label{fig:conv_schematic}
\end{figure*}

In a Monte Carlo simulation, where the interactions or decays would be treated on an event-by-event basis, the direction $\hat{\mathbf{u}}_{h}$ of the secondary particle $h$ in the fixed (lab) frame could be found via a simple addition of $\hat{\mathbf{u}}_{{\ell}}$ and $\hat{\mathbf{u}}_{{\ell} \rightarrow h}$. However, to incorporate angular evolution into the semi-analytical cascade theory, the distributions of the particle travel directions have to be formulated in terms of \textit{angular densities}. Invoking the azimuthal symmetry, i.e., the invariance with respect to $\varphi$, we define the double-differential particle density $\eta$ with respect to the energy and the polar angle as

\begin{equation}
    \eta_{h}(E, \theta) = \frac{1}{\theta}\frac{\mathrm{d}^2 N_h( \theta)}{ \mathrm{d}\theta \mathrm{d}E},
\label{eq:2d_ang_density_definition}
\end{equation}
which is normalized to the single-differential density:
\begin{equation}
    n_h(E) = \int_{0}^{\theta_{ \mathrm{max}}} \eta_h (E, \theta) \theta \mathrm{d}\theta.
\label{eq:flux_norm_definition}
\end{equation}
\noindent Throughout this study, we assume $\theta_{\mathrm{max}} = \pi/2$, i.e.\@ consider only forward-going particles, as well as the delta function-like angular distributions of the primaries. As the cascade develops, more secondaries will be produced off-axis and the angular distribution $\eta_h$ of the secondaries will evolve as a function of slant depth. On the other hand, the distribution of the relative angles between the primaries and the secondaries, $\theta_{l \rightarrow h}$, is defined by the allowed phase space in a given interaction or decay process and constitutes a fixed \textit{convolution kernel}. The angular distribution of the secondary particle $h$ can then be obtained as a \textit{two-dimensional convolution} of the primary angular density with the convolution kernel in the plane orthogonal to the primary particle direction. For the case of secondaries obtained in interactions, we denote this kernel as $\varsigma_{l \to h}$ and write
\begin{subequations}
\begin{align}
    \eta_{h} (E, \theta) &= \eta_{\ell} (E, {\theta}_{\ell}) \ast \ast\,\varsigma_{\ell \to h} ({E_{\ell}, E, \theta}_{\ell \to h}),\\
    & = \int_{0}^{\theta_{\max}} \eta_{\ell} ({\theta_{\ell})} \,\varsigma_{\ell (E_{\ell}, \theta_{\ell}) \rightarrow h (E, \theta)} \theta_{\ell} \mathrm{d} {\theta}_{\ell}, 
\end{align}
\label{eq:twodim_convolution}
\end{subequations}
where the ``$\ast \ast$'' operator represents two-dimensional convolution. The convolution kernels for decays will be denoted as $\delta_{l \to h}$ throughout this study. Following the formalism of \cite{Baddour:2009xxx} for the convolution of two azimuthally symmetric functions, we have absorbed the integration over the azimuthal variables into the definition of $\varsigma_{\ell \rightarrow h}$ and $\delta_{\ell \rightarrow h}$.
To illustrate the 2D convolution principle, we consider as an example a proton-induced hadronic cascade, as shown in \autoref{fig:conv_schematic}. In this simplified setup, a beam of protons with the energy density $n$ enters the atmosphere at the slant depth $X_0$ aligned with the downward-pointing $z$ axis, hence $\theta_{\mathrm{primary}} = 0$ . The direction of this proton beam is represented by the unit vector $\hat{\mathbf{u}}_{\mathrm{primary}}$.
In the 1D geometry, the velocity unit vector $\hat{\mathbf{u}}_{\mathrm{secondary}}$ of $\nu_{\mu}$ is aligned with $\hat{\mathbf{u}}_{\mathrm{primary}}$, while in the 2D geometry, this does not hold beyond $X_0$. As the proton interacts with the atmospheric nuclei at $X_1$, the secondary products of the interaction (including the $\pi^{+}$) gain transverse momentum, and their angular distribution widens. This is represented by the convolution with the kernel $\varsigma_{p \to \pi^{+}}$. The angular distribution of muon neutrinos at $X_2$ further widens due to the convolution of the pion angular density with the decay kernel $\delta_{\pi^{+} \to \nu_{\mu}}$.

Mathematically, the production of the particle $h$ with the energy $E$ by the interactions of the primary $\ell$ with the energy $E_{\ell}$ leads to the following change in the angular density of $h$:

\begin{equation}
    \frac{\mathrm{d}\eta^{\mathrm{int}}_h (\theta)}{\mathrm{d}X} = \frac{1}{\lambda_{\mathrm{int}, \ell}}\int_{0}^{\pi/2} \eta_{\ell}(\theta_{\ell}) \varsigma_{\ell (E_{\ell}, \theta_{\ell}) \rightarrow h (E, \theta)} \theta_{\ell}\,\mathrm{d}\theta_{\ell}.
\label{eq:convolution_operators}
\end{equation}
An equivalent expression can be formulated for decays by replacing $\varsigma_{\ell \to h}$ with $\delta_{\ell \to h}$. The appearance of the $\theta_\ell$ factor in the integrals of \cref{eq:convolution_operators} is an important feature of the 2D convolution in the $xy$ plane, where $\theta_\ell$ and $\theta$ are interpreted as the radii of the $\hat{\mathbf{u}}_{l}$ and $\hat{\mathbf{u}}_h$ velocity vectors projected onto $xy$.\\

\subsection{Two-dimensional cascade equations in the angular domain}\label{sec:analytical_twodim_equations}

\Cref{eq:convolution_operators} and its equivalent for decays are the source terms in the two-dimensional cascade equations; they directly modify the angular densities of the secondaries, which evolve longitudinally as a function of the slant depth $X$ in the atmosphere. At the same time, the sink terms in \cref{eq:ana_onedim_cascade_sinks} and \cref{eq:ana_onedim_energy_losses} do not change the angular distribution of the primaries and only contribute to the change in the overall normalization. With these two observations combined, we can write down the 2D version of \autoref{eq:ana_onedim_cascade_eq} as follows:
\begin{widetext}
\begin{align}
\frac{\mathrm{d} \eta_h(E, X,\theta)}{\mathrm{d}X} = & -\frac{\eta_h(E, X,\theta)}{\lambda_{\mathrm{int}, h}(E)}-\frac{\eta_h(E, X,\theta)}{\lambda_{\mathrm{dec},h} (E, X)} - \frac{\partial}{\partial E} (\mu_E \eta_h(E, X, \theta)) \nonumber \\ &+ \sum_{\ell} \int_{0}^{\pi/2} \theta_{\ell} \mathrm{d}\theta_{\ell} \int_{E}^{\infty} \mathrm{d} E_{\ell} \frac{\varsigma_{\ell (E_{\ell}, \theta_{\ell}) \to h (E, \theta)} }{\lambda_{\mathrm{int},\ell}(E_{\ell})} \eta_{\ell}(E_{\ell}, X,\theta_{\ell})  \label{eq:twodim_cascade_eq} \\&+ \sum_{\ell} \int_{0}^{\pi/2} \theta_{\ell} \mathrm{d}\theta_{\ell} \int_{E}^{\infty} \mathrm{d} E_{\ell} \frac{\delta_{\ell (E_{\ell}, \theta_{\ell}) \to h (E, \theta)} }{\lambda_{\mathrm{dec},\ell}(E_{\ell}, X)} \eta_{\ell}(E_{\ell}, X,\theta_{\ell}) \nonumber.   
\end{align}
\end{widetext}
The longitudinal development of the secondary particle cascades is computed through the forward difference integration of \autoref{eq:twodim_cascade_eq}. The new component in \autoref{eq:twodim_cascade_eq} compared to \autoref{eq:ana_onedim_cascade_eq} is the angular development of the secondaries, which is taken care of via the 2D convolutions of the angular densities of the primaries with the interaction/decay convolution kernels.

\section{Matrix cascade equations and the MCEq code}\label{sec:mceq}

The basic principle of the \textsc{MCEq} code is to evolve the hadronic and electromagnetic cascades in the atmosphere, given a cosmic ray primary flux and the probabilities of interactions and decays of all primary and secondary particles. With these inputs, \textsc{MCEq} solves the cascade equations for the densities of the secondaries of interest. We begin this section with a review of the numerical form of the one-dimensional equations (\autoref{eq:ana_onedim_cascade_eq}) based on the formalism derived in \cite{Fedynitch:2015zma, Fedynitch:2018cbl, Fedynitch:2022vty}. We further extend this numerical framework to two dimensions, building on \cite{Kozynets:2021zcp}.

\subsection{Review of the matrix cascade equations in 1D}\label{sec:1d_equations_matrix}

For a shower particle $h$ which can interact with nuclei in the atmosphere (e.g.\@ $^{14}$N or $^{16}$O), the interaction cross section is energy-dependent, as is the yield of the interaction products in an inelastic collision. Additionally, if the particle is unstable, the energy spectra of its decay products depend on the boost of the parent particle. It is therefore natural to discretize the transport equation in energy, i.e.\@ to represent the particle densities in discrete energy bins $E_i$, $i \in [0, N_E - 1]$. The discrete one-dimensional cascade equation reads:

\begin{subequations}
\begin{align}
    \frac{\mathrm{d} n^h_{E_i}(X)}{\mathrm{d}X} = & -\frac{n^h_{E_i} (X)}{\lambda_{\mathrm{int},E_i}^h} -\frac{n^h_{E_i} (X)}{\lambda_{\mathrm{dec},E_i}^h (X)} \label{eq:onedim_cascade_sinks}\\
    &-\boldsymbol{\nabla}_i [\mu_{E_i}^h n_{E_i}^h (X)]
    \label{eq:onedim_energy_losses}\\
    & + \sum_{\ell} \sum_{E_k^{*} \geq E_i^{*}} \frac{c_{\ell (E_k) \to h (E_i)} }{\lambda^{\ell}_{\mathrm{int},E_k}} n^{\ell}_{E_k} (X)\label{eq:onedim_cascade_inter_sources}\\
    & + \sum_{\ell} \sum_{E_k^{*} \geq E_i^{*}} \frac{d_{{\ell} (E_k) \to h (E_i) }}{\lambda^{\ell}_{\mathrm{dec},E_k} (X) } n^{\ell}_{E_k} (X)\label{eq:onedim_cascade_decay_sources}, 
\end{align}
\label{eq:onedim_cascade_eq}
\end{subequations}
where we arranged the terms in the same order as in \cref{eq:ana_onedim_cascade_eq} to clarify the correspondences between the continuous and the discrete equation versions. In \cref{eq:onedim_cascade_inter_sources}, we defined the coefficient $c$ for the yield of particle $h$ in interactions as
\begin{equation}
    c_{\ell (E_k) \to h (E_i)} = \frac{\mathrm{d}N_{\ell(E_k) \to h(E_i)}}{\mathrm{d}E}\Big|_{E=E_i} \Delta E_k,
\label{eq:1d_yield_coefficient_def}
\end{equation}
which translates as the energy density of particles $h$ with energy $E_i$ generated per primary $\ell$ within the energy bin $E_k$. The decay coefficients $d$ in \cref{eq:onedim_cascade_decay_sources} are defined in the same way. The equations for the different particle species are coupled through the yield coefficients $c_{\ell \to h}$ and $d_{\ell \to h}$. The solution is obtained by solving \cref{eq:onedim_cascade_eq} in $X$ \cite{Fedynitch:2015zma, Fedynitch:2018cbl} iteratively in the matrix form. The yield coefficients are derived from the event generators (e.g.\@ \textsc{UrQMD} \cite{Bass:1998ca}, \textsc{DPMJet} \cite{Roesler:2000he, Fedynitch:2015kcn}, \textsc{Sibyll} \cite{Riehn:2019jet, Fedynitch:2018cbl}, or \textsc{epos-lhc} \cite{Pierog:2013ria} for hadron-nucleus collisions, and \textsc{Pythia} \cite{Bierlich:2022pfr} for decays) by histogramming the secondary particle yields as a function of the secondary particle kinetic energy, $E_i$. In the 1D approximation, all secondary particle angles with respect to the primary particle direction of motion are contributing to the yield coefficient, thereby resulting in an angle-integrated interaction/decay probability. 

\subsection{2D matrix cascade equations in the Hankel frequency domain}\label{sec:hankel_space_equations}

Depending on the energy scales of hadronic interactions and unstable particle decays in the atmosphere, the widths of the angular distributions of the secondary particles can vary by orders of magnitude. For example, the average emission angle of \SI{2}{GeV} pions produced in a collision of a \SI{100}{GeV} is $10^{\circ}$ with respect to the primary proton direction, while \SI{20}{GeV} pions deflect only by $\sim$$1^{\circ}$ from the proton axis and produce $\nu_{\mu}$ at angles as small as $0.01^{\circ}$ relative to the pion direction. Due to the evolution of hadronic cascades over multiple generations (shower age), the total deflection is amplified by the number of generations even if the angular deflection in a single interaction/decay is small. As a result, \autoref{eq:twodim_cascade_eq} requires a ``universal'' $\theta$ grid which could accommodate both large and small angular deflections. Making such a grid linear would imply an extremely fine discretization, and the numerical evaluation of the 2D convolution integrals would become prohibitively expensive. If the $\theta$ grid was logarithmic, the computation of the convolution integral would become more complicated due to the mis-alignment of the input and the output grids. While the techniques for convolving functions defined on logarithmic grids exist, they usually come with hyperparameters to be tuned by the user in order to keep the numerical errors to the minimum \cite{Haines:1988xxx, Hamilton:1999uv, Lang:2019xxx}. This extra freedom in the choice of hyperparameters could lead to unpredictable numerical behaviour in the integration of \autoref{eq:twodim_cascade_eq} over thousands of steps in $X$. 

To avoid the complications of the 2D convolutions in the $\theta$ space (which we will also refer to as the ``real'' space), we choose to operate in the spectral (``frequency'') domain instead. This is motivated by the existence of the \textit{convolution theorem}, which transforms the convolutions in the real space into multiplications in the frequency space. For the specific case of the 2D convolution of the azimuthally symmetric functions $\varsigma_{\ell \to h}, \delta_{\ell \to h}$, and $\eta(X, \theta_{\ell})$, the correct transform enabling the use of the convolution theorem is the zeroth-order Hankel transform $\mathcal{H}$ \cite{Baddour:2009xxx}:
\begin{equation}
    \mathcal{H}[f(\theta)](\kappa) = \int_{0}^{\infty} f(\theta) J_{0} (\kappa\theta)\,\theta\,\mathrm{d}\theta,
\label{eq:hankel_transform_definition}
\end{equation}
where $f(\theta)$ is a function of the continuous variable $\theta$, $\kappa$ is the spectral frequency mode ($\kappa \geq 0$), and $J_0$ is the zeroth-order Bessel function of the first kind. In the formal definition of $\mathcal{H}$, the upper limit of the $\theta$ integral in \autoref{eq:hankel_transform_definition} is $\infty$, however we only consider the forward-going particles with $\theta \leq \pi/2$. 

The convolution theorem states that, for the azimuthally symmetric functions $f(\theta)$ and $g(\theta)$, 
\begin{equation}
   \mathcal{H}[f(\theta) \ast \ast\,g(\theta)] = \mathcal{H} [f(\theta)](\kappa) \cdot \mathcal{H} [g(\theta)](\kappa),
\label{eq:convolution_theorem}
\end{equation}
i.e., the Hankel transform of the convolution result is a product of the Hankel transforms of the input functions in the frequency space \cite{Baddour:2009xxx}.
\Cref{eq:convolution_theorem} is fully applicable to the two-dimensional cascade equations in \cref{eq:twodim_cascade_eq}. We therefore bring the convolution kernels and the angular densities of the cascade particles to the Hankel frequency space by defining their zeroth-order Hankel transforms as follows:
\begin{subequations}
\begin{align}
   &\tilde{\eta}^h_{E_i} (X, \kappa) \equiv \mathcal{H} [{\eta}^h_{E_i} (X, \theta)] (\kappa); \\
   &\tilde{\varsigma}_{\ell(E_k) \to h(E_i)} (\kappa) \equiv \mathcal{H} [{\varsigma}_{\ell(E_k) \to h(E_i)} (\theta)] (\kappa);\\
   &\tilde{\delta}_{\ell(E_k) \to h(E_i)} (\kappa) \equiv \mathcal{H} [{\delta}_{\ell(E_k) \to h(E_i)} (\theta)] (\kappa).
\end{align}
\label{eq:hankel_transforms_of_kernels_and_fluxes}
\end{subequations}
Then, we can reformulate \autoref{eq:twodim_cascade_eq} as

\begin{align}
\frac{\mathrm{d} \tilde{\eta}^h_{E_i}(X, \kappa)}{\mathrm{d}X} = & -\frac{\tilde{\eta}^h_{E_i} (X, \kappa)}{\lambda_{\mathrm{int},E_i}^h}-\frac{\tilde{\eta}^h_{E_i} (X, \kappa)}{\lambda_{\mathrm{dec},E_i}^h (X)} \nonumber \\& -\boldsymbol{\nabla}_i [\mu_{E_i}^h \tilde{\eta}^h_{E_i}(X, \kappa)] \nonumber 
\\&+\sum_{E_k^{*} \geq E_i^{*}} \sum_{\ell}  \frac{[\tilde{\varsigma}_{{\ell}(E_k) \to h(E_i)} \cdot \tilde{\eta}_{E_k}^l] (\kappa)} {\lambda^{\ell}_{\mathrm{int},E_k}}\label{eq:hankel_space_cascade_eq}\\ &+\sum_{E_k^{*} \geq E_i^{*}} \sum_{\ell} \frac{[\tilde{\delta}_{{\ell}(E_k) \to h(E_i)} \cdot \tilde{\eta}_{E_k}^{\ell}] (\kappa)}{\lambda^{\ell}_{\mathrm{dec},E_k} (X)}\nonumber, 
\end{align}
which is the main equation to be solved in ``2D \textsc{MCEq}''.

For practical applications, the $\kappa$ grid is made discrete and integer-valued. This implies that in \autoref{eq:hankel_space_cascade_eq}, the multiplication of the Hankel-transformed convolution kernels and the Hankel-transformed angular densities of the primaries is performed elementwise with respect to the discrete frequency modes $\kappa$. The 1D \textsc{MCEq} equation (\autoref{eq:onedim_cascade_eq}) is a special case of \autoref{eq:hankel_space_cascade_eq} for $\kappa = 0$, as $J_0(0) = 1$ and \autoref{eq:hankel_transform_definition} becomes equivalent to our earlier definition of the angular density normalization from \autoref{eq:flux_norm_definition}. Therefore,  \autoref{eq:hankel_space_cascade_eq} retains the computational complexity of \autoref{eq:onedim_cascade_eq}, up to a linear scaling by the number of the frequency modes ($N_{\kappa}$). One can then either choose to solve the $N_{\kappa}$ equations (one for each $\kappa$) sequentially or in parallel, or to assemble the Hankel-transformed yield coefficients and angular densities into a more complex matrix structure. Our current implementation relies on the sequential solution of the $N_{\kappa}$ equations but can easily be adapted to the user's preference. In \cref{sec:event_generation,sec:2d_matrices}, we explain how to arrive at the Hankel-transformed yield coefficients for the relevant interaction and decay channels, as well as provide further details on our choice of the $\kappa$ grid where these coefficients are stored.

\section{2D MCEq pipeline and solution scheme}\label{sec:2d_mceq_pipeline}

The main computational advantage of the \textsc{MCEq} code compared to the Monte Carlo simulations comes from the pre-calculation of the particle yields, which is done outside of the user interface. The pre-tabulated interaction and decay coefficients are used to build the matrices for \autoref{eq:onedim_cascade_eq} (1D \textsc{MCEq}) or \autoref{eq:hankel_space_cascade_eq} (2D \textsc{MCEq}) during the code initialization.
These coefficients are derived from the histogrammed kinematic properties of the secondary particles in hadronic interactions or decays, as evaluated via the \textsc{chromo} code \cite{CHROMO:2022xxx}. The histograms for 1D \textsc{MCEq} include primary and secondary particle energies ($E_{\mathrm{prim}}$ and $E_{\mathrm{sec}}$). For 2D \textsc{MCEq}, an additional Hankel mode $\kappa$ is incorporated as a third dimension, which stores the angular densities of the secondaries in a compact form. Below, we describe how the output of any given event generator is used to populate the $(E_{\mathrm{prim}}, E_{\mathrm{sec}}, \kappa)$ grid.

\subsection{Compact representation of event information} \label{sec:event_generation}

For the low-energy atmospheric neutrino flux calculations ($\mathcal{O}$(GeV) and below), we deploy a logarithmically-spaced kinetic energy grid ranging from \SI{10}{MeV} to \SI{10}{TeV}\footnote{2D \textsc{MCEq} solutions for atmospheric neutrino fluxes are numerically stable down to energies around \SI{50}{MeV}.}. This grid utilizes a bin width of $\Delta\log_{10} E_{\mathrm{kin}} = 0.1$, resulting in $N_E = 60$ energy bins. The 2D \textsc{MCEq} code currently excludes electromagnetic cascades, deemed unnecessary for the evolution of hadronic cascades that produce $\mathcal{O}$(GeV) atmospheric leptons\footnote{ Based on our \textsc{corsika} simulations, electromagnetic interactions (such as the muon pair production and the photonuclear interactions) in proton-induced air showers with energies up to \SI{1}{TeV} contribute $\lesssim$1-2\% to the energy density of atmospheric leptons. The full cosmic ray spectrum follows an inverse power law, and the contribution of cosmic rays with energies $\gg$\SI{1}{TeV} to the $\mathcal{O}$(GeV) atmospheric lepton flux is suppressed.}. Thus, we apply \autoref{eq:hankel_space_cascade_eq} only to hadrons and leptons (excluding the $\tau$ lepton), totaling $H = 21$ particle species: 6 baryons ($p/\bar{p}, n/\bar{n}$, and $\Lambda^{0}/\bar{\Lambda}^{0}$), 5 mesons ($\pi^{\pm},\,K^{\pm},\,K_{\mathrm{L}}^0,$), and 10 leptons ($\mu^{\pm}_{\mathrm{R/L}},\,\mu^{\pm}, \,\nu_{e}/\bar{\nu}_e$, and $\nu_{\mu}/\bar{\nu}_{\mu}$). Muons contribute 6 species, where each $\mu^{\pm}$ includes two polarizations: left-handed ``L'', right-handed ``R'', and an unpolarized component (denoted as $\mu^{\pm}$ without a subscript). 

For hadronic interactions, the \textsc{chromo} code \cite{CHROMO:2022xxx} runs the following models: \textsc{UrQMD} \cite{Bass:1998ca}, \textsc{epos-lhc} \cite{Pierog:2013ria}, \textsc{Sibyll-2.3d} \cite{Fedynitch:2018cbl, Riehn:2019jet}, and \textsc{DPMJet-III 19.1} \cite{Roesler:2000he, Fedynitch:2015kcn}. \textsc{Pythia 8.306} \cite{Bierlich:2022pfr} is used for unstable particle decays. However, it cannot simulate the production of polarized muons in $\pi^{\pm}$ and $K^{\pm}$ two-body decays or the three-body decays of polarized muons, instead generating events in the spin-averaged phase spaces. Hence, muon polarization modeling is done separately, as outlined in \autoref{sec:muon_summary} and detailed in \cref{appendix:muon_details}.

Our event generation and histogramming scheme is consistent across all interaction/decay channels. For every primary in the kinetic energy bin $k$, we assign the logarithmic center ($E_k = \sqrt{E_{k'} \cdot E_{k''}}$ for $[E_{k'}, E_{k''}]$) as the primary's initial energy. This primary particle enters the chosen event generator with four-momentum $p^{\mu}_{\mathrm{prim}} = (E_k, 0, 0, \sqrt{E_k^2 + 2 E_k m_{\mathrm{prim}}})^{\top}$, moving along the positive $z$ axis. A stationary nitrogen nucleus ($^{14}$N) is the target for simulating hadronic interactions. Using other atmospheric nuclei like $^{16}$O has little effect on secondary particle yields. For decays, we set the unstable particle at rest and boost its decay products to the lab frame, recording the daughter energies as in 1D \textsc{MCEq}. 

To solve the 2D cascade equation (\autoref{eq:hankel_space_cascade_eq}), we additionally need to compute the Hankel-transformed angular densities of secondary particles, i.e., $\tilde{\varsigma}_{l(E_k) \to h(E_i)}(\kappa)$. We illustrate this process using the muon neutrino production chain from \cref{fig:conv_schematic}, i.e., $p + ^{14}\mkern-1.mu\relax\mathrm{N} \to \pi^{+} + X^{*} \to \nu_{\mu} + \mu^{+}$, where $X^{*}$ denotes all other secondary particles and nuclear remnants. As angular distributions of all particles fill a discrete energy grid, we select the primary proton, secondary pion, and tertiary neutrino energy bins around \SI{100}{GeV}, \SI{10}{GeV}, and \SI{4}{GeV}, respectively. This illustrative choice reflects the characteristic energies for the low-energy neutrino production in air showers.

\begin{figure*}[htb!]
\includegraphics[width=\textwidth]{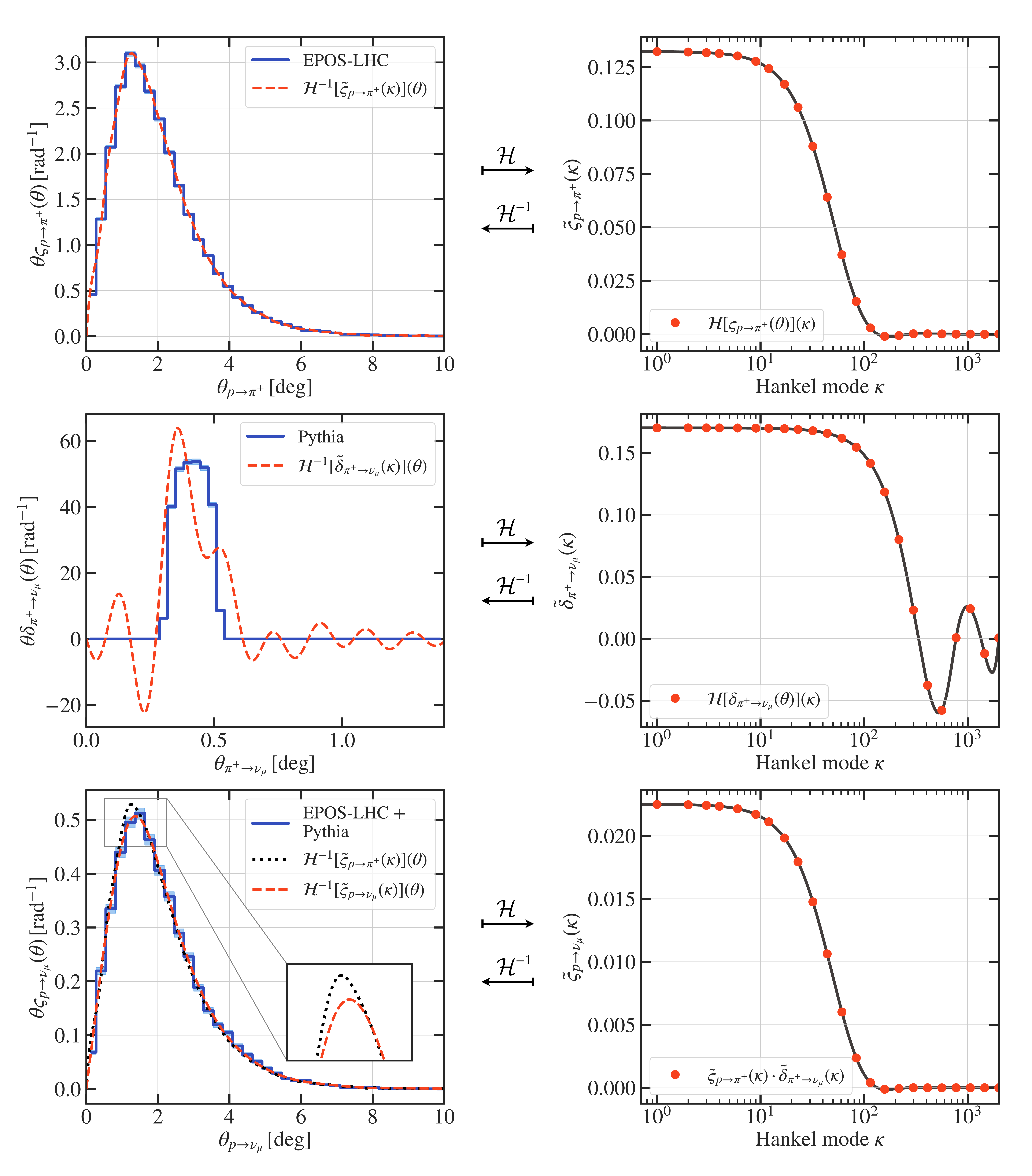}
\caption{\textit{Top left}: Angular distribution of secondary pions in $p + ^{14}$N$ \to \pi^{+} + X^{}$, derived from the \textsc{epos-lhc} event generator (solid blue line), and the inverse Hankel transform result (dashed red line) of the top right panel. \textit{Top right}: Hankel transform of the secondary pions' angular distribution, calculated using \autoref{eq:bessel_function_sum}. \textit{Middle}: Similar to the top panels but evaluated for daughter $\nu_{\mu}$ in $\pi^{+} \to \mu^{+} + \nu_{\mu}$ decay. \textit{Bottom}: Demonstrating the convolution theorem on the $p \to \pi^{+} \to {\nu_{\mu}}$ chain: the inverse transform of the product of the Hankel transforms (top right and middle right) reproduces the angular density of the tertiary neutrinos (bottom left). The dotted black line in the bottom left panel indicates the properly normalized angular distribution of secondary pions for comparison. Energy settings for all considered particles are found in \autoref{tab:energy_bin_settings}. \\}
\label{fig:hankel_chain}
\end{figure*}

The simulation chain begins with protons incident on $^{14}\mkern-1.mu\relax\mathrm{N}$. We choose the \textsc{epos-lhc} hadronic model and compute the yield of the secondary $\pi^{+}$.  The top left panel of \cref{fig:hankel_chain} shows the distribution of angles $\theta_{p \to \pi^{+}}$ that the secondary pions make with the primary proton axis. The number of entries in the histogram, $n_{\pi^{+}} \equiv n_p \cdot c_{p \to \pi^{+}}$, is equal to the total secondary pion yield in this interaction. Each pion contributes a delta function \underbar{$\delta$}$(\theta - \theta_{p \to \pi^{+}}^{j}), j \in [0, n_{\pi^{+}} - 1],$ to the angular density. The Hankel transform of the delta function has an analytical representation, $\mathcal{H}\left[\frac{1}{a}\underbar{$\delta$}(\theta - a)\right](\kappa) = J_0 (\kappa a)$\footnote{Note that the delta function, represented as \underbar{$\delta$}, is distinguished from the decay coefficient $\delta_{l \to h}$ used earlier in \autoref{eq:twodim_cascade_eq}. The $(1 / a)$ scaling of the delta function centered at $\theta=a$ maintains consistency with the angular density definition in \autoref{eq:flux_norm_definition}.}, which can populate the $\kappa$-grid immediately after the event generation. Summation of these Hankel-transformed delta functions approaches the Hankel transform of the secondary pions' underlying angular density. This density can be expressed via inverse Hankel transform as:

\begin{subequations}
\begin{align}
    \varsigma_{p \to \pi^{+}}(\theta) &=  \frac{1}{n_p}\mathcal{H}^{-1}\left[\tilde{\varsigma}_{p \to \pi^{+}}(\kappa)\right],\\
    & = \frac{1}{n_p}\mathcal{H}^{-1}\left[\sum_{j=1}^{n_{\pi^{+}}}J_0 (\kappa \theta_{p \to \pi^{+}}^{j})\right](\theta). 
\end{align}
\label{eq:bessel_function_sum}
\end{subequations}

Equations \ref{eq:hankel_space_cascade_eq} and \ref{eq:bessel_function_sum} could theoretically extend to an infinite number of modes $\kappa$. In practice, a truncated $\kappa$-grid with 24 logarithmically spaced integer modes between 0 and 2000 suffices to accurately represent the angular distributions of GeV-scale atmospheric leptons. In the example shown in \autoref{fig:hankel_chain}, two key observations validate this approach. Firstly, the inverse Hankel transform from \autoref{eq:bessel_function_sum} effectively represents the original pion angular distribution, thus demonstrating the utility of the Hankel transform for compacting angular densities of secondary particles. Secondly, the amplitudes of the higher-frequency modes with $\kappa \geq 100$ are negligible, indicating a sufficiently broad angular distribution of the GeV-scale pions produced in proton-$^{14}$N interactions. For sharper-edged distributions, like in the pion decay to muon neutrinos (middle panel of \autoref{fig:hankel_chain}), the truncated $\kappa$ grid may not sufficiently reconstruct the angular density, resulting in a characteristic ``ringing''. However, in a realistic air shower, the effect of this artefact is minimal due to the wider pion angular distribution.

The final $\nu_{\mu}$ angular distribution from the Monte Carlo simulation chain is successfully reconstructed through the inverse transform of the convolution result in the Hankel space:
\begin{equation}
    \varsigma_{p \to \nu_{\mu}}(\theta) = \mathcal{H}^{-1}\left[\tilde{\varsigma}_{p \to \pi^{+}} (\kappa) \cdot \tilde{\delta}_{\pi^{+} \to \nu_{\mu}(\kappa)}\right](\theta).
\label{eq:conv_theorem_p_pi_numu}
\end{equation}
This demonstrates the application of the convolution theorem from \cref{eq:convolution_theorem}. Notably, after successive convolutions, the final $\theta$ refers to the angle of the secondary particle with respect to the original shower-inducing primary.

\subsection{2D \textsc{MCEq} matrix production}\label{sec:2d_matrices}

We generate 10 million events per primary species and energy bin on the \textsc{MCEq} grid, following the methodology in \autoref{sec:event_generation}. Since different hadronic models have different valid energy ranges, we create interpolated matrices which allow to smoothly transition between the hadronic model choices. This emulates the approach used in codes such as \textsc{corsika} and is explained in \cref{appendix:interpolation}.

The generation of secondary particle yields involves $H=21$ primary particles in 2D \textsc{MCEq}. The yield of the secondaries is stored as a function of the secondary and primary kinetic energies (on a grid of length $N_E$) and the Hankel frequency mode $\kappa$. The resulting dimension of the 2D \textsc{MCEq} matrices is $N_\kappa$  $\times$  ($N_E \cdot H$) $\times$ ($N_E \cdot H$), equating to $24 \times (1260 \times 1260)$. These are treated as $24$ separate sparse matrices when solving the 2D cascade equation in the Hankel frequency domain.

\subsection{Solution in the Hankel space}\label{sec:hankel_solution}

The convolution theorem transforms the 2D convolution into multiplication in the Hankel frequency domain. The amplitude of the primary angular distribution corresponding to the mode $\kappa$ is multiplied by the amplitude of the interaction/decay kernel corresponding to exactly the same mode, i.e., the modes $\kappa_1$ and $\kappa_2$ are not coupled if $\kappa_1 \neq \kappa_2$.  This allows us to treat each of the $N_{\kappa} = 24$ equations of the 2D \textsc{MCEq} completely independently and solve them using the strategy analogous to that of the 1D \textsc{MCEq} \cite{Fedynitch:2018cbl, Fedynitch:2015zma}. This involves applying the forward Euler integrator to \autoref{eq:hankel_space_cascade_eq}, i.e., performing the longitudinal evolution of the Hankel-transformed angular densities of the cascade secondaries in discrete slant depth steps $\Delta X$. This approach is best summarized in the matrix form:
\begin{equation}
\begin{split}
    \tilde{\boldsymbol{\eta}}(X_{t+1}, \kappa) &= \tilde{\boldsymbol{\eta}}(X_{t}, \kappa) - \boldsymbol{\nabla}_E [\mathrm{diag} (\boldsymbol{\mu}) \cdot \tilde{\boldsymbol{\eta}}(X_t, \kappa) ] \\ &+ \Delta X_t \left[ \vphantom{\frac{1}{\rho(X_t)}} (-\mathbb{I} + \mathbfcal{C}_k) \boldsymbol{\Lambda}_{\mathrm{int}} \right. \\
    &+ \left. \frac{1}{\rho(X_t)} (-\mathbb{I} + \mathbfcal{D}_k) \boldsymbol{\Lambda}_{\mathrm{dec}} \right]  \tilde{\boldsymbol{\eta}}(X_{t}, \kappa),
\label{eq:forward_stepping}
\end{split}
\end{equation}
where $X_t$ and $X_{t+1} = X_t + \Delta X_t$ are two consecutive slant depth values, and $\mathbfcal{C}_{\kappa}$ and $\mathbfcal{D}_{\kappa}$ are the slices of the yield coefficient ``cubes'' $\varsigma_{l \to h} (\kappa)$ and $\delta_{l \to h} (\kappa)$ at the frequency mode $\kappa$. Following \cite{Fedynitch:2018cbl, Fedynitch:2015zma}, we also construct the diagonal matrices $\mathbf{\Lambda}_{\mathrm{int}}$ and $\mathbf{\Lambda}_{\mathrm{dec}}$ from interaction and decay lengths. Each diagonal entry corresponds to a specific particle $h$ and energy bin $E_i$, arranged similarly to the particle density vector $\tilde{\boldsymbol{\eta}}(X, \kappa)$ that we seek to evolve. After $\tilde{\boldsymbol{\eta}}(X_{\mathrm{final}}, \kappa)$ is computed, the final step is to reconstruct the angular densities of ${\eta}(X_{\mathrm{final}}, \theta)$ via the inverse Hankel transform.

\subsection{Reconstruction of the real-space solutions}\label{sec:solution_reconstruction}

After the final integration step, the 2D \textsc{MCEq} solver returns the state vector $\tilde{\boldsymbol{\eta}}(X_{\mathrm{final}}, \kappa)$. This includes the Hankel frequency space amplitudes for all cascade particles across the \textsc{MCEq} energy grid. The inverse Hankel transform enables the reconstruction of the secondary particle densities as a function of the angle $\theta$ relative to the primary particle axis:
\begin{subequations}
\begin{align}
    \boldsymbol{\eta}(X_{\mathrm{final}}, \theta) &= \mathcal{H}^{-1} \left[\tilde{\boldsymbol{\eta}}(X_{\mathrm{final}}, \kappa)\right] (\theta), \\ &\equiv \int_{0}^{\infty}\tilde{\boldsymbol{\eta}}(X_{\mathrm{final}}, \kappa) J_0 (\kappa\theta) \kappa\,\mathrm{d}\kappa,\\
    &\simeq \int_{0}^{\kappa_{\mathrm{max}}}\tilde{\boldsymbol{\eta}}(X_{\mathrm{final}}, \kappa) J_0 (\kappa\theta) \kappa\,\mathrm{d}\kappa.\label{eq:inverse_integral_approximation} 
\end{align}
\label{eq:inverse_hankel_transform}
\end{subequations}
Although the $\kappa$ grid in 2D \textsc{MCEq} is discrete with logarithmically spaced modes, accurate computation of the integral in \autoref{eq:inverse_integral_approximation} requires a quasicontinuous $\kappa$ range, which is achieved via spline interpolation of $\tilde{\boldsymbol{\eta}}(X_{\mathrm{final}}, \kappa)$. This can be done at any $X_t$ ($0 \leq X_t \leq X_{\mathrm{final}}$) along the integration path, should the solution at a particular slant depth/altitude be of interest to the user.

The reconstruction of angular densities of high-energy secondaries (with kinetic energies of \SI{10}{GeV} and above), as well as those created very early in the cascade evolution, must be treated with care if the starting angular distribution of the primaries is narrow (e.g.\@ delta function-like). Direct application of \autoref{eq:inverse_hankel_transform} may lead to ``ringing'' in the reconstructed angular densities for such secondary particles. Thus, it is recommended to apply \autoref{eq:inverse_hankel_transform} for secondary particles with energy $\lesssim 10\,\mathrm{GeV}$ at several kilometers into the atmosphere, and to use the 1D approximation at high energies/altitudes. 

\subsection{Modeling of muon transport}
\label{sec:muon_summary}

Muons play a crucial role in air shower development and atmospheric neutrino flux calculations. They either decay in-flight, generating muon and electron (anti)neutrinos, or survive until the surface level, creating a source of background in neutrino detection.

Muon polarization occurs as atmospheric muons are produced in the two-body decays of $\pi^{\pm}$ and $K^{\pm}$. In the parent meson rest frame, the muons are fully polarized, with their momenta perfectly aligned or antialigned with their spin direction. This affects the angular distributions and the energy spectra of neutrinos originating from muon decays \cite{Lipari:1993hd}. A simplified representation of the muon population across the entire continuous range of helicities is achieved by including only six muon species into the 2D \textsc{MCEq} cascade equations (see \autoref{sec:event_generation}).

Another phenomenon affecting muon transport is multiple scattering, which modifies the trajectories of the muons as they scatter with atmospheric nuclei. We implement a Gaussian approximation to the Molière formalism for describing this effect \cite{Heck:1998vt,Bethe:1953va}. This results in an overall widening of the muon angular distributions compared to the case without multiple scattering.

\Cref{appendix:muon_details} details the implementation of muon polarization and multiple scattering, as well as their impact on the results of 2D \textsc{MCEq}.

\section{Benchmarking against CORSIKA}\label{sec:benchmarking}

\subsection{Simulation setup}

To validate our solutions to the two-dimensional matrix cascade equations via 2D \textsc{MCEq}, we use the \textsc{corsika} v7.75 Monte Carlo code \cite{Heck:1998vt} as a benchmark. We aim to compare the angular distributions of the GeV-scale atmospheric neutrinos and muons generated in the cosmic-ray induced air showers. All of our simulations are run for a single angle of incidence of the cosmic ray primary, and the secondary particle angular distributions are computed with respect to the primary particle axis. In the terminology of \autoref{fig:conv_schematic}, we are comparing the distributions of $\arccos(\hat{\mathbf{u}}_{\mathrm{primary}} \cdot \hat{\mathbf{u}}_{\mathrm{secondary}})$ between 2D \textsc{MCEq} and \textsc{corsika}.
 
To ensure a fair comparison, we equalize the configuration settings between \textsc{MCEq} and \textsc{corsika} to the extent possible. Most importantly, we match the choice of hadronic interaction models by using \textsc{UrQMD} \cite{Bass:1998ca} as the low-energy model\footnote{\textsc{corsika} uses the older \textsc{UrQMD-1.3}, while our 2D \textsc{MCEq} matrices were produced with the newer \textsc{UrQMD-3.4} model accessed via a preliminary version of the \textsc{chromo} tool \cite{CHROMO:2022xxx}.} and \textsc{epos-lhc} \cite{Pierog:2013ria} as the high-energy model in both \textsc{corsika} and 2D \textsc{MCEq}. The transition energy between models is set to \SI{150}{GeV}\footnote{The \textsc{UrQMD} model was tested and shown to give reasonable results in the energy range $E_{\mathrm{lab}} = \,$\SIrange[range-phrase=--,range-units=single]{2}{160}{GeV} \cite{Steinheimer:2009nn}. The usage of the model up to the maximum energy of the Relativistic Heavy Ion Collider (\SI{250}{GeV} for proton beams) is considered valid \cite{urqmd_user_guide}.}. In \cref{appendix:cm_benchmarking_details}, we provide a comprehensive list of other relevant physics settings in \textsc{MCEq} and \textsc{corsika}.

Our setup for the benchmarking of lepton densities consists of a proton primary incident onto the Earth's atmosphere at an inclination angle $\theta_0$ with respect to the negative $z$ axis (downward direction). We test both vertical ($\theta_0 = 0^{\circ}$) and inclined showers ($30^{\circ} \leq \theta_0 \leq \theta_{\mathrm{max}} = 80^{\circ}$). The energy of the proton either is fixed or follows a spectrum with a power-law dependence (e.g.\@ $\propto E^{-2.7}$). For each considered initial condition, we simulate $\sim$1 million events in \textsc{corsika} with different random seeds. This lets us gather enough statistics for the low-energy muons and neutrinos at several observation altitudes. The corresponding binned angular distributions are compared directly to the angular densities obtained with 2D \textsc{MCEq} by solving \autoref{eq:hankel_space_cascade_eq}. 

\subsection{Results}\label{sec:corsika_comparisons}

\begin{figure*}[htpb]
  \includegraphics[width=\textwidth]{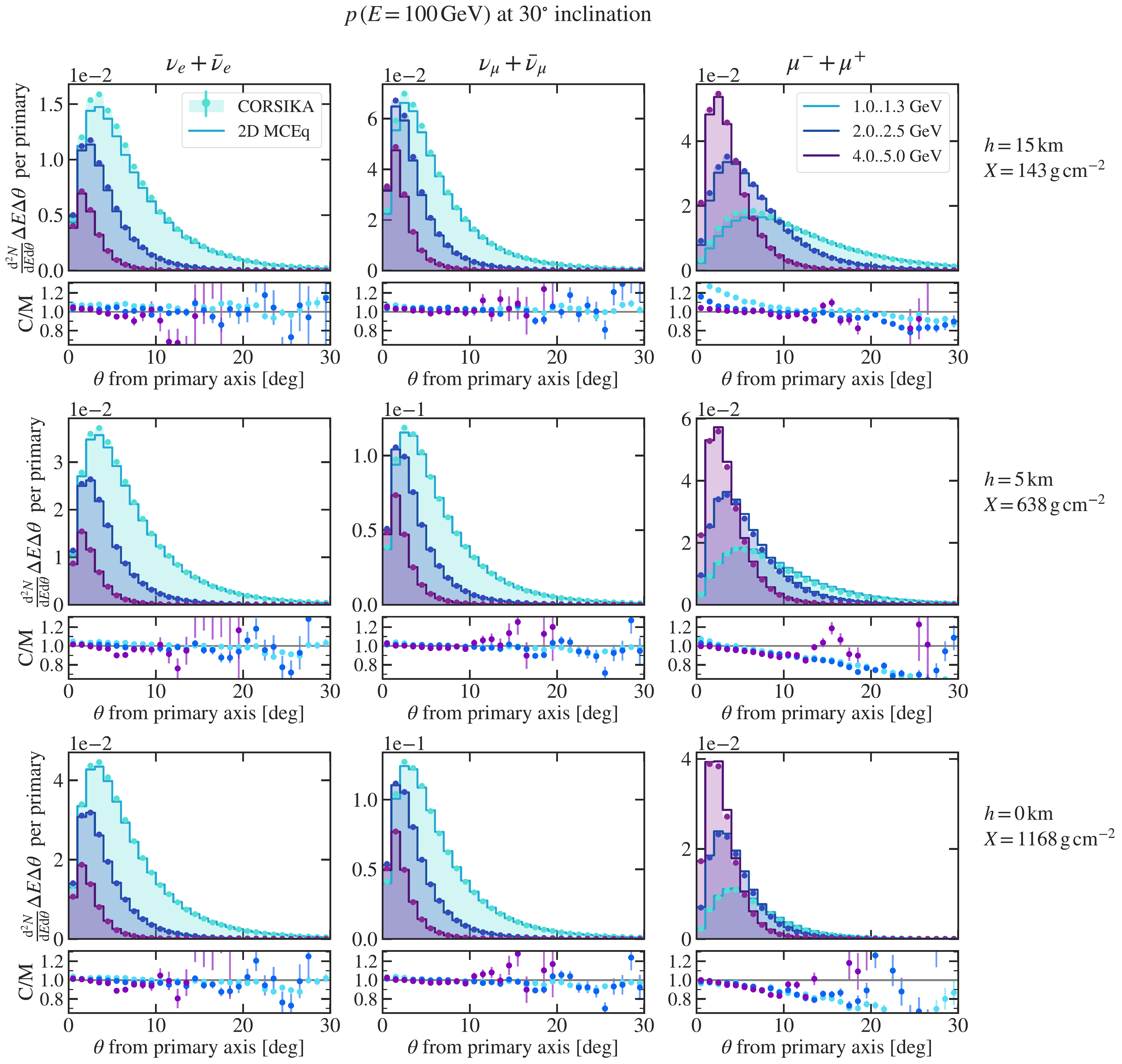}
  \caption{Angular distributions of atmospheric leptons in a proton-induced air shower ($E_0$ = \SI{100}{GeV}, $\theta_0 = 30^{\circ}$), as computed numerically in 2D \textsc{MCEq} (solid lines) and simulated in the \textsc{corsika} Monte Carlo (filled histograms with errorbars). The angle $\theta$ on the $x$ axis is the angle a given secondary makes with the direction of the primary proton. The different colors correspond to the different energy bands, and the bottom sub-panel in each plot shows the ratio of \textsc{corsika} (``C'') to \textsc{MCEq} (``M'').}
  \label{fig:ang_dists_100GeV_30deg}
\end{figure*}

To provide a representative example of how the 2D \textsc{MCEq} solutions compare to the \textsc{corsika} Monte Carlo outputs, we choose the case of a \SI{100}{GeV} proton primary incident at $\theta_0 = 30^{\circ}$. In \autoref{fig:ang_dists_100GeV_30deg}, we show the angular densities of neutrinos ($\nu_{\mu} + \bar{\nu}_{\mu}$; $\nu_{e} + \bar{\nu}_{e}$) and muons ($\mu^{-}$ + $\mu^{+}$) at low energies (up to \SI{5}{GeV}).

\begin{figure*}
  \includegraphics[width=1\textwidth]{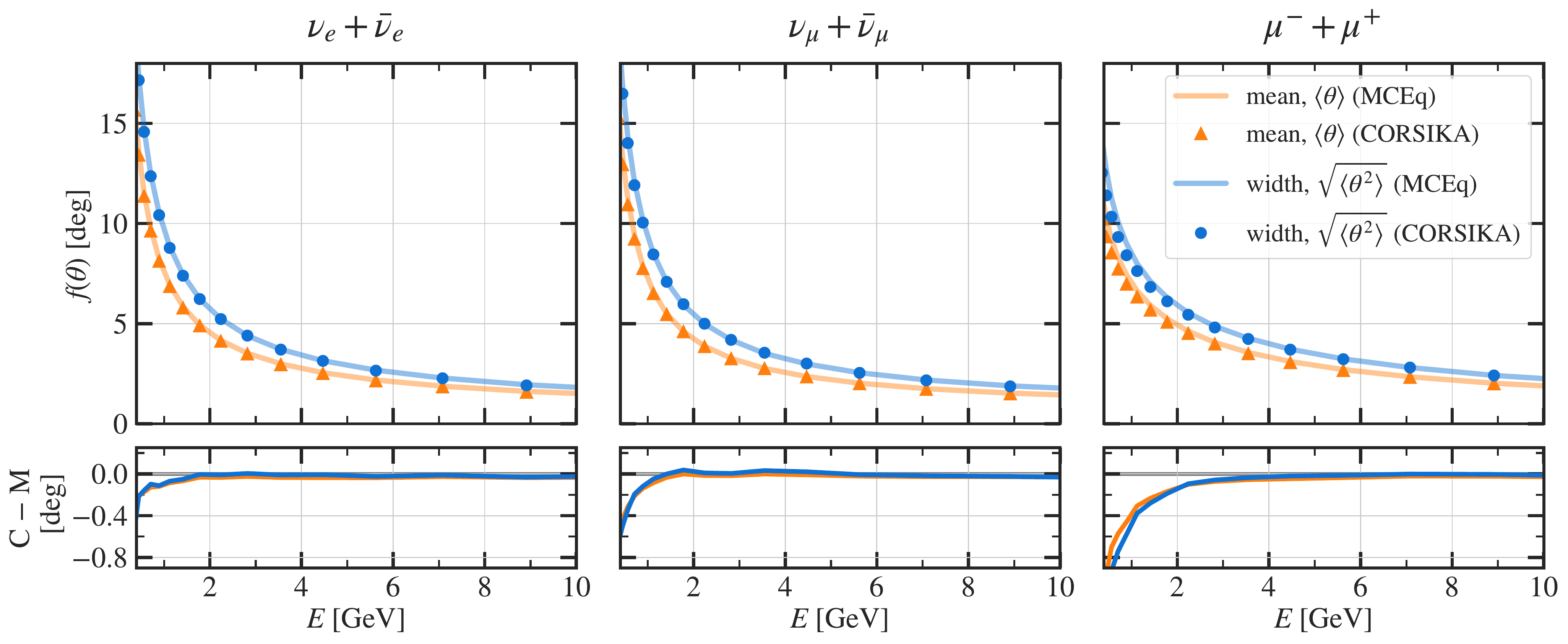}
  \caption{Comparison of the first ($\langle \theta \rangle$, or the distribution mean) and the second ($\sqrt{\langle \theta^2 \rangle}$, or the distribution width) moments of the angular distributions of atmospheric leptons at the Earth's surface as computed in \textsc{corsika} (``C'') and \textsc{MCEq} (``M'') for the same initial conditions as in \cref{fig:ang_dists_100GeV_30deg}. The bottom panel shows the difference between the \textsc{corsika} and the \textsc{MCEq} estimates.}
  \label{fig:angular_moments_100GeV_30deg}
\end{figure*}

From \autoref{fig:ang_dists_100GeV_30deg}, we find that the angular distributions of $\mathcal{O}$(few GeV) leptons with respect to the proton primary axis are in a very good agreement between 2D \textsc{MCEq} and \textsc{corsika}. For neutrinos, the differences between the two codes are predominantly statistical and reach at most $10\%$ in the tails of the distributions. This level of agreement holds across all altitudes and energy bins considered. For muons, a characteristic tilt of the \textsc{corsika}-to-\textsc{MCEq} angular distribution ratio is observed at all altitudes, reaching $\sim$20$\%$ in the distribution tails. This is indicative of a bias of the \textsc{corsika} angular distribution towards smaller angles or the \textsc{MCEq} angular distributions towards larger angles. One possible explanation for this pattern is that all particles in 2D \textsc{MCEq} travel exactly the same distance to reach a specific depth, including those that deflect by as much as \SIrange[range-phrase=--,range-units=single]{20}{30}{}$^{\circ}$ from the primary axis. However, muons with such large deflection angles naturally travel longer distances than those at $0^{\circ}$, introducing a factor of $1 / \cos \theta$ increase in the integration step length. This means that muons with large deflections from the primary direction must lose more energy than currently modelled in 2D \textsc{MCEq} and migrate from the energy bins shown in \autoref{fig:ang_dists_100GeV_30deg} to the bins of lower energy. Qualitatively, this explains why \textsc{corsika} could have fewer muons at large angles. However, since the amount of energy lost is directly proportional to the distance travelled $(\Delta E \simeq \langle \frac{\mathrm{d}E}{\mathrm{d}X}\rangle \, \Delta X$), any discrepancies related to the energy loss are expected to accumulate with distance. This shows to a small degree in \autoref{fig:ang_dists_100GeV_30deg}, where the tilt in the \textsc{corsika}-to-\textsc{MCEq} ratio of the muon angular densities develops mildly as a function of altitude. While we cannot definitively connect the angle-dependent discrepancy between 2D \textsc{MCEq} and \textsc{corsika} with the simplified treatment of the angle-dependent propagation step length in 2D \textsc{MCEq}, the latter remains a relevant feature to be implemented in future iterations of the \textsc{MCEq} code. At present, we point out that the 2D \textsc{MCEq} angular densities still provide a very good overall representation of the \textsc{corsika} angular distributions, which can be seen from the agreement of the distribution moments ($\langle \theta \rangle$ and $\langle \theta^2 \rangle$) in \autoref{fig:angular_moments_100GeV_30deg}. The sub-degree level of difference observed in the angular distribution moments will not be possible to resolve under any realistic experimental resolution at GeV-scale energies, implying a negligible impact on experimental analyses.

\begin{figure*}[htpb]
  \includegraphics[width=\textwidth]{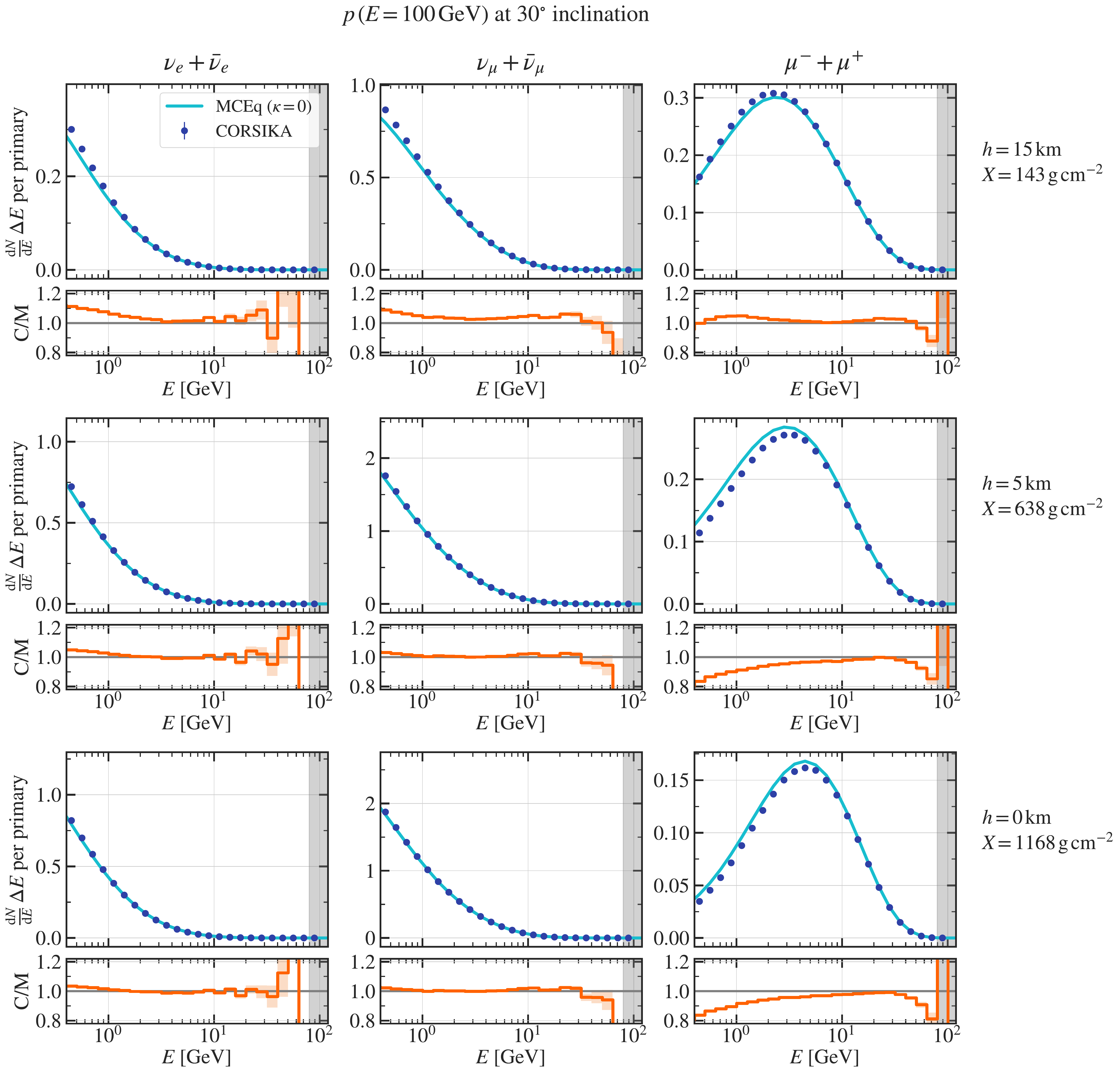}
  \caption{Energy spectra of atmospheric leptons in a proton-induced air shower ($E_{0}$ = \SI{100}{GeV}, $\theta_0 = 30^{\circ}$), as computed numerically in 1\textsc{MCEq} (solid lines) and simulated in the \textsc{corsika} Monte Carlo (filled markers). Here, ``\textsc{MCEq}'' corresponds to the $\kappa = 0$ slice of the 2D \textsc{MCEq} solution. The bottom sub-panel in each plot shows the ratio of \textsc{corsika} (``C'') to \textsc{MCEq} (``M''). The shaded gray band represents the region where the \textsc{MCEq} solution is numerically unstable (see the caption of \autoref{fig:polarized_muon_impact_spectra} for details).}
  \label{fig:en_spectra_100GeV_30deg}
\end{figure*}

While the main objective of the 2D \textsc{MCEq} code is to evolve angular distributions of the secondaries in addition to the energy spectra already provided by 1D \textsc{MCEq}, we compare the energy spectra from 2D \textsc{MCEq} to those from \textsc{corsika} to provide further validation to our approach. To obtain the energy spectra from 2D \textsc{MCEq}, we extract the $\kappa = 0$ mode from the Hankel-space solutions, which is equivalent to the angle-integrated particle densities as per \autoref{sec:hankel_space_equations}. As seen in \autoref{fig:en_spectra_100GeV_30deg}, the spectra from \textsc{MCEq} and \textsc{corsika} agree within a few $\%$ in the \SIrange[range-phrase=--,range-units=single]{1}{10}{GeV} region, which is the main energy range of interest in this study. Above \SI{10}{GeV}, the difference between the two codes grows as a function of energy, reaching $\sim$20$\%$ at the maximum neutrino energies available from \SI{100}{GeV} primary showers. The energy dependence of the \textsc{corsika}-to-\textsc{MCEq} ratio could point to the difference in the treatment of hadronic interactions, e.g.\@ the hadron yields between the different \textsc{UrQMD} model versions. The same level of agreement is observed when  the primaries have power law-like spectra, as demonstrated in \cref{fig:ang_dists_spectrum_30deg,fig:en_spectra_spectrum_30deg} in \cref{appendix:power_law_spectrum}.

For shower inclinations less than $60^{\circ}$, our tests demonstrate the same level of agreement as in \cref{fig:ang_dists_100GeV_30deg,fig:en_spectra_100GeV_30deg}. In \cref{fig:ang_dists_100GeV_80deg,fig:en_spectra_100GeV_80deg} in \cref{appendix:horizontal_showers}, a comparison between 2D \textsc{MCEq} and \textsc{corsika} is further presented for highly inclined showers (80$^{\circ}$), revealing up to $25\%$ differences in the angular distributions at large angles of deflection from the primary particle axis. While it is possible that the different implementations of muon energy losses or muon propagation geometries are contributing to the observed mismatch, the precise impact of these factors has not been quantified in this study. However, we emphasize that the level of agreement of the angular distributions and spectra in the \SIrange[range-phrase=--,range-units=single]{1}{10}{GeV} energy region is still satisfactory even for highly inclined showers, and the mild angle-dependent discrepancy observed for single-primary showers will be smeared out by the integration of the secondary particle fluxes across the full sky.

Finally, we note that both the 2D \textsc{MCEq} solutions and the \textsc{corsika} Monte Carlo outputs are subject to systematic uncertainties due to the choice of the hadronic interaction model used to describe the particle yields in the hadron-nucleus inelastic collisions. For the energy spectra and angular distributions of the low-energy leptons, the choice of the low-energy hadronic interaction model ($E_{\mathrm{primary}} \leq 150\,\mathrm{GeV}$) has the largest impact. We test two hadronic interaction models internally within 2D \textsc{MCEq} in \cref{fig:ang_dists_le_model_comparison,fig:en_spectra_le_model_comparison} in \cref{appendix:impact_of_le_hadronic_models}, finding up to 20\% differences between the models at lepton energies below \SI{10}{GeV}.

\section{Summary and outlook}\label{sec:discussion}

In this work, we have detailed the development and application of 2D \textsc{MCEq}, an extended version of the 1D \textsc{MCEq} software. The 2D \textsc{MCEq} code provides an efficient numerical approach to angular evolution of hadronic cascades with broad particle physics applications -- in particular, in atmospheric lepton flux modelling. This tool considers all crucial aspects of hadronic and leptonic physics, such as inelastic interactions of hadrons with atmospheric nuclei, decays of unstable particles, energy losses, muon polarization, and muon multiple scattering.

Validation of 2D \textsc{MCEq} was performed against the standard Monte Carlo code, \textsc{corsika}. The results display agreement within \SIrange[range-phrase=--,range-units=single]{1}{10}{\%} for neutrino angular distributions in air showers up to medium inclinations. Larger differences between the two codes were observed in the distribution tails (corresponding to large angles and high energies).

Given the very high level of agreement with \textsc{corsika} and a significant computational superiority over the Monte Carlo approach, 2D \textsc{MCEq} provides a very appealing option for atmospheric lepton flux calculations. The computational cost of the 2D \textsc{MCEq} calculations is between several CPU-seconds for vertical showers and 1 CPU-minute for the near-horizontal showers, compared to multiple CPU-hours to gather sufficient statistics for inclusive flux calculations via the Monte Carlo simulations. Our tool therefore opens the pathway to fast exploration of the systematic uncertainties on the angular distributions of atmospheric leptons, including those associated with the hadronic interaction models and the cosmic ray primary flux. The 2D \textsc{MCEq} code can further be utilized within hybrid air-shower calculation frameworks, such as the integration of \textsc{corsika} and \textsc{conex} \cite{Bergmann:2006yz}, with the added feature of explicit angular dependence. 

Future enhancements will involve the integration of three-dimensional calculations, accounting for factors such as the Earth's spherical geometry, the initial angular distribution of cosmic ray primaries, the geomagnetic cutoff for these primaries, and the deflection of cascade secondaries within the geomagnetic field.

\section*{Acknowledgements}
The authors, T.K. and D.J.K., acknowledge the support from the Carlsberg Foundation (project no. 117238). The author, A.F., is grateful for the supportive environment provided by Prof.~Hiroyuki Sagawa's group at the ICRR as a recipient of the JSPS International Research Fellowship (JSPS KAKENHI Grant Number 19F19750). The authors acknowledge the invaluable computational resources provided by the Academia Sinica Grid-Computing Center (ASGC), which is supported by Academia Sinica. 

\clearpage
\appendix

\counterwithin{figure}{section}
\setcounter{table}{0}
\renewcommand{\thetable}{A\arabic{table}}

\onecolumngrid

\section{Details of the MCEq matrix production}\label{appendix:technical_details_matrix_production}

\subsection{Event generation chain example}\label{appendix:event_generation_details}

\begin{table*}[htb!]
    \centering
    \begin{tabular}{lllllll}
    \hline\\[-1em]
        particle && $\log_{10}\left(\frac{E_{k'}}{\mathrm{GeV}}\right)\,[E_{k'}\,(\mathrm{GeV})]$ & \quad & $\log_{10}\left(\frac{E_{k}}{\mathrm{GeV}}\right)\,[E_{k}\,(\mathrm{GeV})]$ &\quad  & $\log_{10}\left(\frac{E_{k''}}{\mathrm{GeV}}\right)\,[E_{k''}\,(\mathrm{GeV})]$ \\ \hline
        $p$ && 2.0 [100.0] & \quad & 2.05 [112.2] & \quad & 2.1 [125.8] \\ \hline
        $\pi^{+}$ && 1.0 [10.0] &\quad & 1.05 [11.2] & \quad & 1.1 [12.6] \\ \hline
        $\nu_{\mu}$ && 0.5 [3.2] &\quad & 0.55 [3.5] & \quad & 0.6 [4.0]  \\ \hline        
    \end{tabular}
    \caption{Energy grid settings for the event generation chain example in \cref{fig:hankel_chain}. The three columns (from left to right) correspond to the left bin edge, the bin center, and the right bin edge of the respective particles on the \textsc{MCEq} kinetic energy grid.}
    \label{tab:energy_bin_settings}
\end{table*}

\subsection{Hadronic model interpolation}\label{appendix:interpolation}

\begin{figure}[h!]
  \includegraphics[width=8cm]{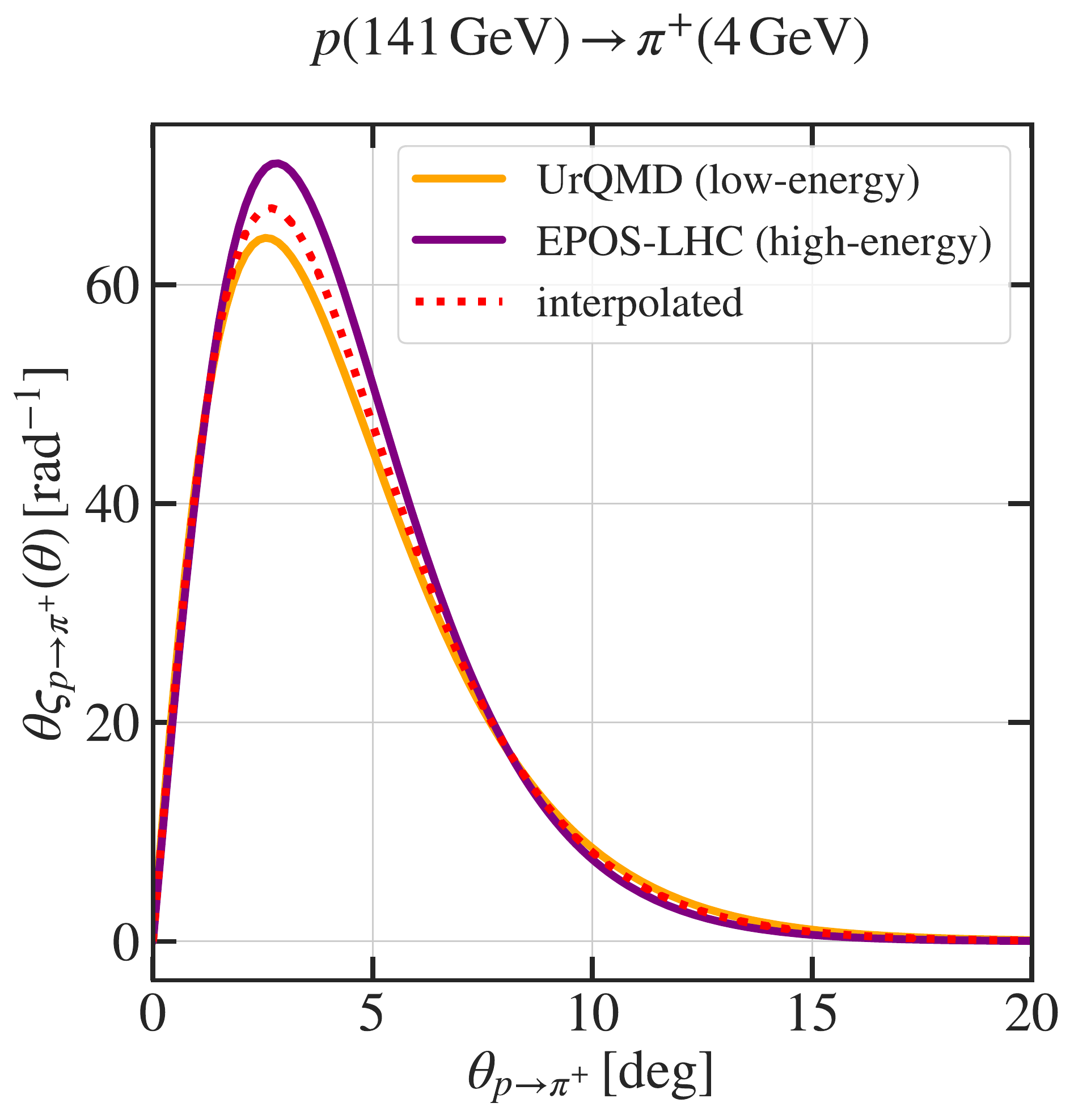}
  \caption{Angular density of the secondary pions obtained in the $p + ^{14}$N$ \to \pi^{+} + X^{*}$ process, as simulated in the \textsc{UrQMD} (orange) and the \textsc{epos-lhc} event generators. Assuming the threshold of \SI{150}{GeV} in the low-/high-energy model transition, the protons in the \textsc{MCEq} energy bin centered at $\sim$\SI{140}{GeV} fall into the ``intermediate'' energy regime, where we use a model linearly interpolated between \textsc{UrQMD} and \textsc{epos-lhc}.}
  \label{fig:model_interpolation_in_2d}
\end{figure}

The \textsc{Sibyll-2.3d} and \textsc{epos-lhc} generators are the common choice for the high-energy hadronic interaction model in the air shower codes such as \textsc{corsika}. They are nominally valid down to the primary energies of $E_{\mathrm{thresh}} = 80\,\mathrm{GeV}$ \cite{Heck:1998vt}. At $E_{\mathrm{primary}} < E_{\mathrm{thresh}}$, the \textsc{UrQMD} or \textsc{DPMJet-III 19.1} models are recommended. In our tests, the transition threshold is lifted to $E_{\mathrm{thresh}} = 150\,\mathrm{GeV}$, where the low-energy \textsc{UrQMD} model is still valid \cite{urqmd_user_guide,Steinheimer:2009nn}.
To transition between the two energy regimes, we run both ``high’’ and ``low’’ energy models in \textsc{chromo} and interpolate across the two energy bins adjacent to $E_{\mathrm{thresh}}$ (one on each side of $E_{\mathrm{thresh}}$). In this case, ``linear spline'' refers to the order-1 spline interpolation on the logarithmic \textsc{MCEq} energy grid. An example result of the interpolated angular distributions of the secondary pions in the $p + ^{14}$N$ \to \pi^{+} + X^{*}$ process is shown in \autoref{fig:model_interpolation_in_2d}.

\subsection{Muon production and propagation}\label{appendix:muon_details}

Along with the generic hadronic cascade development, the cascade equations need to account for muon-specific phenomena -- namely, \textit{muon polarization} and \textit{muon multiple scattering} -- for accurate GeV and sub-GeV neutrino flux predictions. The implementation of these phenomena in 2D \textsc{MCEq} is outlined below.

\subsubsection{Muon polarization}\label{sec:muon_polarization_implementation}

Most atmospheric muons are produced in the two-body decays of $\pi^{\pm}$ and $K^{\pm}$. In the rest frame of the decaying mesons, the muons are completely polarized, i.e.\@ have their momenta perfectly aligned or antialigned with their spin direction. This is a direct consequence of the angular momentum conservation. For example, $\mu^{-}$ always has a negative helicity (projection of the muon spin onto its momentum) in the $\pi^{-}/K^{-}$ rest frame, since $\bar{\nu}_{\mu}$ generated in the same decay must necessarily be right-handed (i.e., have a positive helicity). A similar argument holds for $\mu^{+}$ and $\nu_{\mu}$, with the helicity assignments flipped. In the lab frame, the $\mu^{-}$ helicity in the two-body decay of a meson $\mathcal{M^{-}}$ reads \cite{Lipari:1993hd}:
\begin{equation}
\mathcal{P}(\beta_{\mathcal{M}}, \theta^{*}) = \frac{1}{\beta_{\mu}} \cdot \frac{(1 - r_{\mathcal{M}}) + (1 + r_{\mathcal{M}}) \cos \theta^{*} \beta_{\mathcal{M}}}{(1 + r_{\mathcal{M}}) + (1 - r_{\mathcal{M}}) \cos \theta^{*} \beta_{\mathcal{M}}},
\label{eq:muon_helicity}
\end{equation}
where $r_{\mathcal{M}} = ({m_{\mu}} / {m_{
\mathcal{M}}})^2$, $\beta$ is the Lorentz velocity factor ($\beta_X \equiv v_X / c$), and $\theta^{*}$ is the angle of the muon emission in the decaying meson rest frame (defined e.g.\@ with respect to the $z$ axis). For each individual muon, the helicity defined in \autoref{eq:muon_helicity} is a continuous quantity spanning the range between -1..1. To simplify the solution of the cascade equations, where the helicity expectation values of the population of muons rather than the spin states of individual muons are of interest, we switch to the \textit{helicity basis}, i.e.\@ the basis of purely right-handed muons $\mu^{\pm}_{\mathrm{R}}$ and purely left-handed muons $\mu^{\pm}_{\mathrm{L}}$. Then, on average, the probability of finding $\mu^{-}$ in the right-/left-handed state is
\begin{equation}
    \mathcal{P}_{\mathrm{R, L}}(\beta_{\mathcal{M}}, \theta^{*}) = \frac{1}{2} \left[1  \pm \mathcal{P} (\beta_{\mathcal{M}}, \theta^{*})\right]
\label{eq:helicity_to_probability},
\end{equation}
where ``+'' corresponds to the right-handed state and ``-'' -- to the left-handed state. For $\mu^{+}$, the correspondence between the right-/left-handedness and the +/- sign in \autoref{eq:helicity_to_probability} is flipped. In practice, to find the probabilities of the polarized muon production, we let $\pi^{\pm}$ and $K^{\pm}$ decay at rest in \textsc{Pythia} and select the two-body decay events. This immediately gives us the muon emission angle $\theta^{*}$ and, after boosting to the lab frame, the velocity factors $\beta_{\mu}$ and $\beta_{\mathcal{M}}$. Then, we compute the probabilities of the right-/left-handedness according to \autoref{eq:helicity_to_probability} and assign the respective helicity to each generated muon. We also keep track of the muon lab frame energies and their emission angles with respect to the primary meson boost axis. Finally, we fill in the ``cubes'' of the polarized muon yield coefficients in the Hankel frequency space as outlined before in \autoref{sec:event_generation}. 

To compute and pre-histogram the neutrino yields in the decays of polarized muons, we use the \textsc{whizard} Monte Carlo code \cite{Kilian:2007gr, Moretti:2001zz}, which can take the spin direction of the parent muon as an input and sample from the final three-particle phase space simultaneously. This is in contrast with the analytical expressions for the daughter lepton momenta in the polarized muon decay provided in \cite{Lipari:1993hd} and \cite{Mine:1996xx}, which are given separately for each daughter and marginalized over momenta of the other two decay products. Thus, we prefer the \textsc{whizard} Monte Carlo approach for simplicity and efficiency of implementation. For each of $\mu^{-}$ and $\mu^{+}$ decaying at rest, and each of the two choices of spin configurations ($\hat{\boldsymbol{s}}_{\mu} = \langle 0, 0, \pm 1 \rangle$), we generate 10,000,000 three-body decay events. We then boost the daughter neutrinos to the muon energies corresponding to the kinetic energy grid of \textsc{MCEq}. This gives us the angular distributions and the energy spectra of neutrinos originating from the decays of the left-helical and the right-helical muon states $\mu^{\pm}_{\mathrm{L}}$ and $\mu^{\pm}_{\mathrm{R}}$. The superposition of the latter, as prescribed by \autoref{eq:helicity_to_probability}, gives an accurate representation of the muon population across the entire continuous range of helicities accessible to the polarized muons via \autoref{eq:muon_helicity}. This justifies the inclusion of only six muon species ($\mu^{\pm}_{\mathrm{L}}$, $\mu^{\pm}_{\mathrm{R}}$, as well as unpolarized muons $\mu^{\pm}$ originating from the three-body decays of $K^{\pm}$) into the \textsc{MCEq} cascade equations.

\begin{figure}
  \includegraphics[width=8cm]{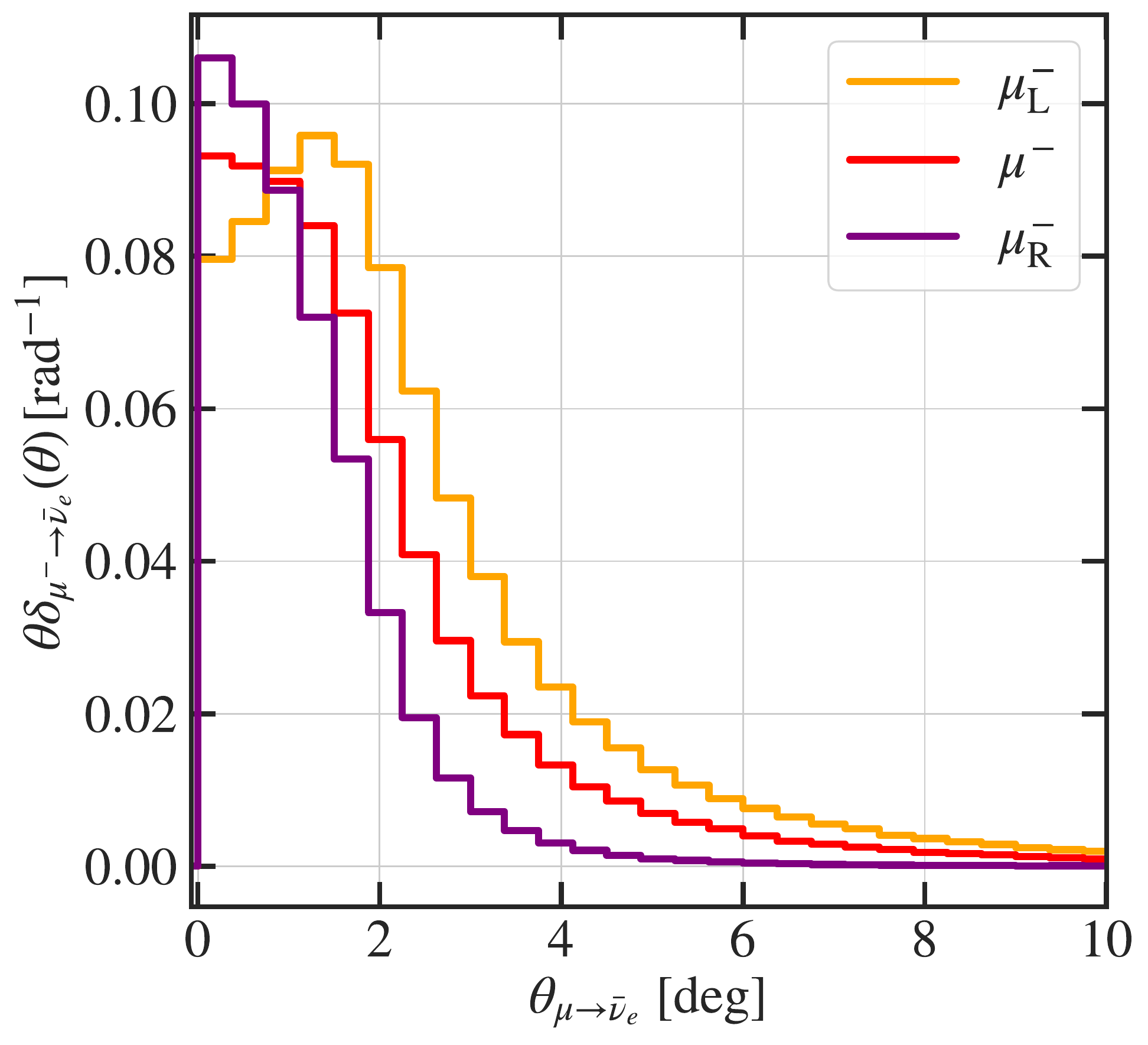}
  \caption{Comparison of the $\bar{\nu}_{e}$ angular distributions with $E_{\bar{\nu}_{e}} \leq 2\,\mathrm{GeV}$ in the decay of a \SI{5}{GeV} muon for three muon polarization cases: left-handed (yellow), right-handed (purple), and unpolarized (red). The displayed events were generated with the \textsc{whizard} Monte Carlo code \cite{Kilian:2007gr, Moretti:2001zz}.}
  \label{fig:polarized_muon_decay}
\end{figure}

In \autoref{fig:polarized_muon_decay}, we show example angular distributions the electron antineutrinos resulting from the $\mu^{-} \to \nu_{\mu} + \bar{\nu}_e + e^{-}$ decay as computed in \textsc{whizard}. For example, at $E_{\mu} = 5\,\mathrm{GeV}$ and $E_{\bar{\nu}_{e}} \leq 2\,\mathrm{GeV}$, we find that both the shape of the neutrino angular distributions and their normalization differs depending on muon polarization. While \cite{Fedynitch:2015zma, Fedynitch:2018cbl} already included the muon polarization effects in the one-dimensional approximation via the analytical prescription of \cite{Lipari:1993hd}, this simple example further illustrates the importance of the muon polarization treatment for the two-dimensional solver.

\begin{figure*}
  \includegraphics[width=0.85\textwidth]{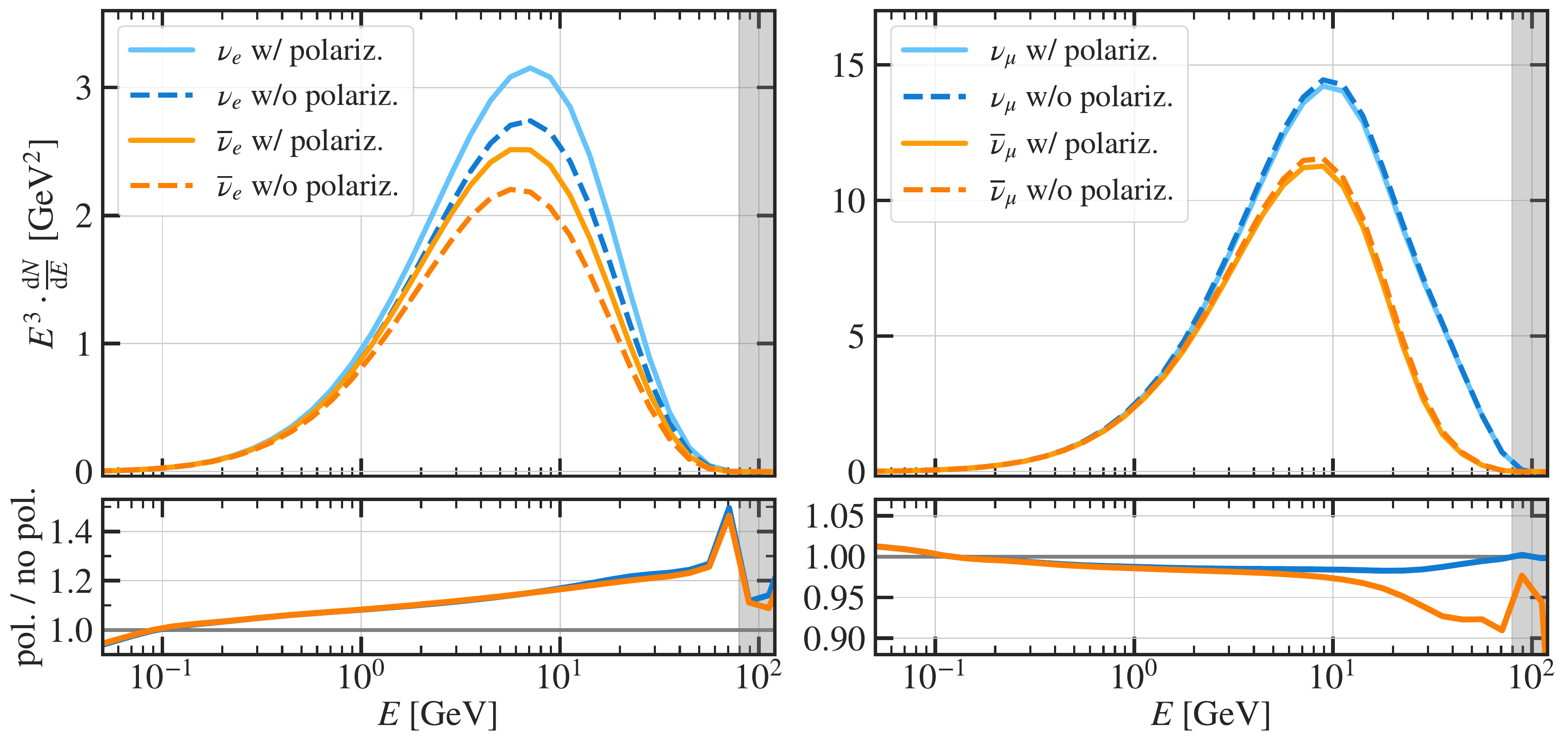}
  \caption{Impact of muon polarization on the energy spectra of the sea-level atmospheric neutrino fluxes in a \SI{100}{GeV} proton air shower with $30^{\circ}$ inclination. The dashed lines refer to the case of all atmospheric muons being treated as unpolarized. The solid lines correspond to the muon polarization treatment as described in the main text of the present section.  The shaded gray band represents the region where the \textsc{MCEq} solution is unstable due to the numerical implementation of the delta function-like initial condition on the discrete \textsc{MCEq} energy grid (see \cite{Heinze:2019jou} for the discretization details). }
  \label{fig:polarized_muon_impact_spectra}
\end{figure*}

\begin{figure*}
  \includegraphics[width=0.87\textwidth]{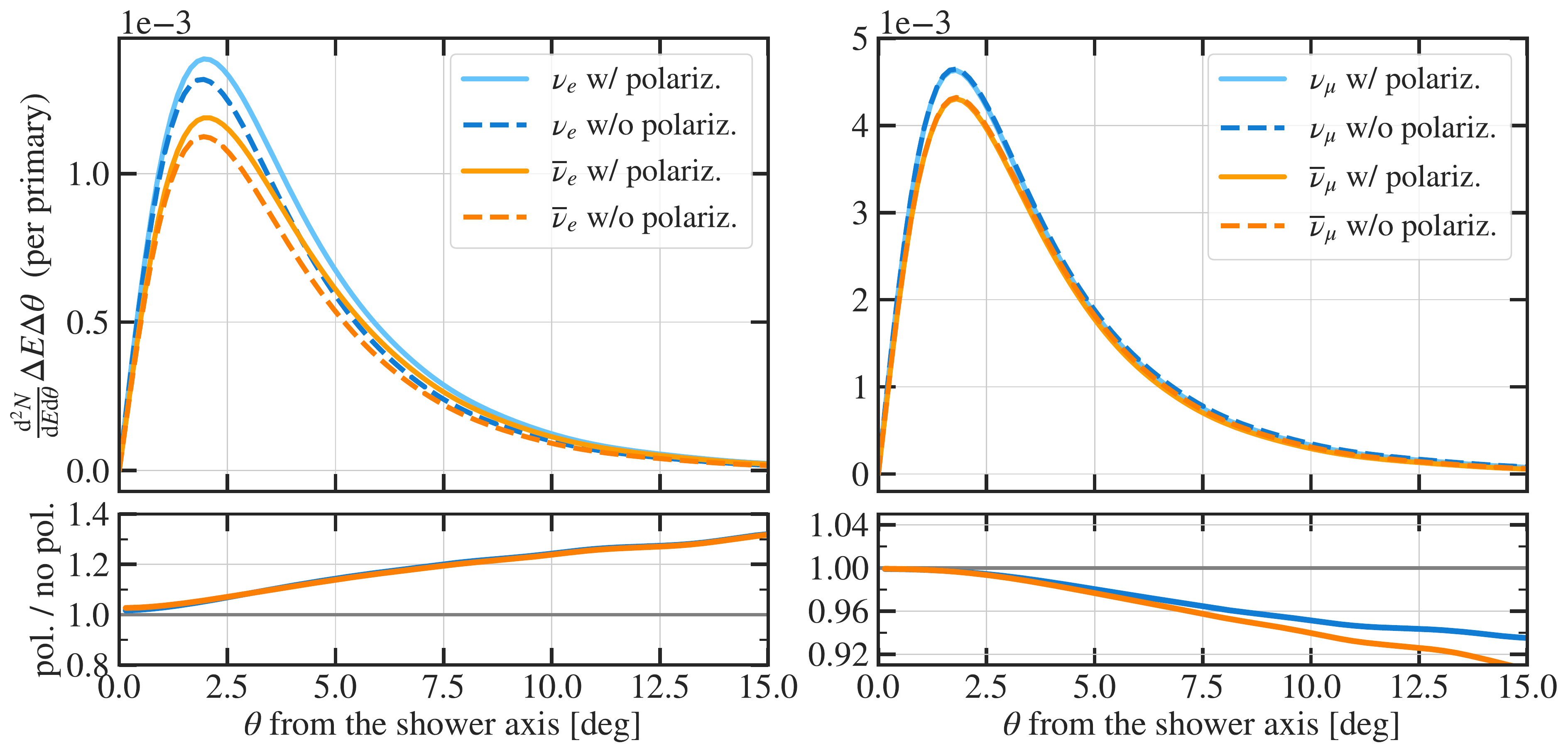}
  \caption{Impact of muon polarization on the angular distributions of the sea-level atmospheric neutrinos with $E_{\nu} \simeq 2\,\mathrm{GeV}$. The initial conditions of the air shower are the same as in \autoref{fig:polarized_muon_impact_spectra}.}
  \label{fig:polarized_muon_impact_angdists}
\end{figure*}

In \cref{fig:polarized_muon_impact_spectra,fig:polarized_muon_impact_angdists}, we show the impact of the muon polarization on the energy spectra and angular distributions of atmospheric neutrinos produced in a full proton-induced air shower. In this example, the primary proton has a fixed energy of \SI{100}{GeV} and a $30^{\circ}$ inclination. The neutrino fluxes with muon polarization enabled/disabled are obtained via 2D \textsc{MCEq} as the solutions to \autoref{eq:hankel_space_cascade_eq} at the sea level. This corresponds to $X \approx 1196\,\mathrm{g\,cm^{-2}}$ in the US Standard atmosphere \cite{Heck:1998vt}. We find that muon polarization has the largest impact on the $\nu_e$ fluxes and energy spectra. Assuming that all muons are unpolarized can lead to 10-30$\%$ error in the $\nu_e$ spectrum normalization and the nearly the same bias in the angular density. The spectrum and the angular distribution of $\nu_{\mu}$ are affected at the level of a few percent; $\bar{\nu}_{\mu}$ experience an up 10$\%$ effect growing towards higher energies due to the energy-dependent ${\pi^{+}}/\pi^{-}$ and ${\mu^{+}}/\mu^{-}$ ratios. The muon normalizations and angular distributions remain unchanged as expected.

\subsubsection{Muon multiple scattering}

\begin{figure}
  \includegraphics[width=8cm]{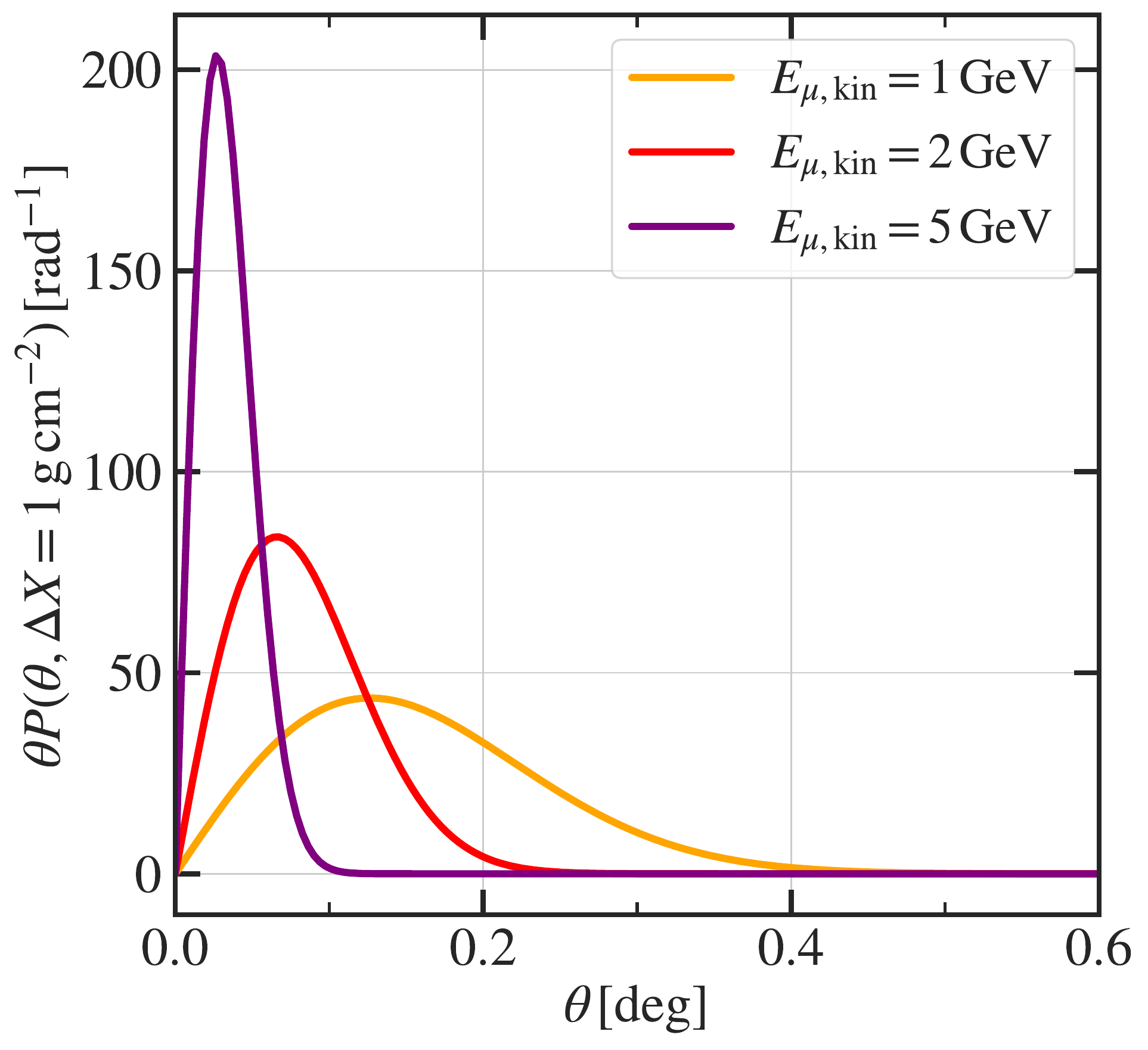}
  \caption{Gaussian approximation of the muon deflection angles due to their multiple scattering on atmospheric nuclei (see \cite{Heck:1998vt, Groom:2000xx, Bethe:1953va} and \autoref{eq:muon_multiple_scattering_angle}). The assumed slant depth traversed by the muons is $\Delta X = 1\,\mathrm{g\,cm^{-2}}$. As expected, lower-energy muons (e.g.\@ \SI{1}{GeV}; yellow line) get deflected more than the higher-energy muons (e.g.\@ red and purple lines for \SI{2}{GeV} and \SI{5}{GeV}, respectively).} 
  \label{fig:muon_mult_scat_kernels}
\end{figure}

An additional effect modifying the muon angular distributions is their Coulomb scattering with atmospheric nuclei, e.g.\@ $^{14}$N or $^{16}$O. The effect of multiple scattering is described by the Molière theory \cite{Bethe:1953va} if the number of scatters is large ($\gg 20$ across a layer of matter with a given thickness), and by the Poissonian probability of scattering with the Rutherford cross section if the number of scatters is small \cite{Heck:1998vt, Groom:2000xx}. We implement a Gaussian approximation to the Molière formalism, which is one of the multiple scattering handling options provided in the \textsc{corsika} Monte Carlo \cite{Heck:1998vt}. In the Gaussian approximation, the probability $P$ of a muon deflecting by the space (unprojected) angle $\theta$ after traversing $\Delta X$ of the atmospheric slant depth is defined as follows:
\begin{equation}
P(\theta, \Delta X) = \frac{1}{{\pi \theta_{\mathrm{s}}^2 \Delta X}} \cdot \exp\left[\frac{-\theta^2}{\Delta X \theta_{\mathrm{s}}^2}\right],
\label{eq:muon_multiple_scattering_angle}
\end{equation}
where $\theta_{\mathrm{s}}^2 = \frac{1}{\lambda_{\mathrm{s}}} \left( \frac{E_{\mathrm{s}}}{E_{\mu, \mathrm{lab}}\, \beta_{\mu}}\right)^2$, $E_{\mathrm{s}} = 0.021\,\mathrm{GeV}$, $\lambda_{\mathrm{s}} = 37.7\,\mathrm{g\,cm^{-2}}$, $E_{\mu, \mathrm{lab}}$ is the total muon energy in the lab frame, and $\beta_{\mu}$ -- its lab-frame Lorentz velocity factor \cite{Heck:1998vt, Groom:2000xx, Rossi:1952xx}. In \autoref{fig:muon_mult_scat_kernels}, we show several representative angular densities computed according to \autoref{eq:muon_multiple_scattering_angle}. For illustration, we use $\Delta X = 1\,\mathrm{g\,cm^{-2}}$. In general, however, $\Delta X$ varies with $X$ in response to the longitudinal atmospheric density variations, and the width of the muon multiple scattering kernel is variable. While in just $1\,\mathrm{g\,cm^{-2}}$ the expected muon deflection is small ($\mathcal{O}(0.1^{\circ})$ at GeV energies), this effect accumulates with the slant depth and results in a noticeable shift of the sea-level muon angular distribution, especially in horizontal showers.

To incorporate this additional convolution kernel into the solution of \autoref{eq:hankel_space_cascade_eq}, we need to Hankel-transform \autoref{eq:muon_multiple_scattering_angle}. In the Hankel frequency space, the muon multiple scattering kernel reads:
\begin{equation}
    \tilde{P}(\kappa, \Delta X) = \exp\left[{\frac{-\kappa^2 \Delta X \theta_{\mathrm{s}}^2}{4}}\right],
\label{eq:muon_multiple_scattering_hankel}
\end{equation}
which is scaled so that the overall normalization of the muon angular distribution (represented by the $\kappa = 0$ mode) remains unchanged. We can then directly multiply \autoref{eq:muon_multiple_scattering_hankel} by the Hankel amplitudes of the muon angular distributions after each integration step $\Delta X$. This way, the simplified muon multiple scattering model becomes a natural part of the matrix cascade equations. The treatment of multiple scattering is identical for all of the muon species, i.e.\@ $\mu^{\pm}_{\mathrm{L}}, \mu^{\pm}_{\mathrm{R}}$, and $\mu^{\pm}$.

\begin{figure}
  \includegraphics[width=8cm]{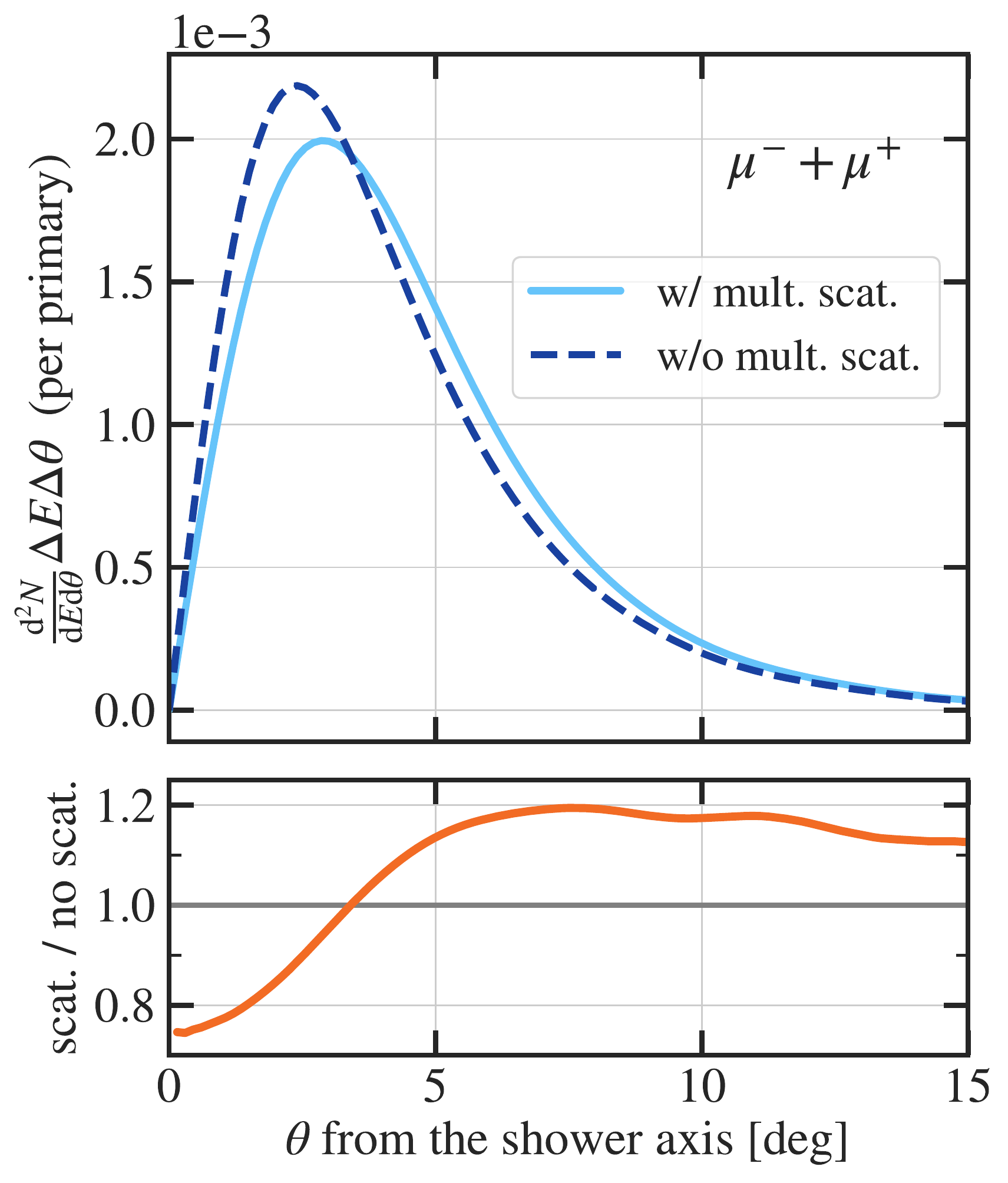}
  \caption{Impact of muon multiple scattering on the angular distributions of the sea-level atmospheric muons with $E_{\mu} \simeq 2\,\mathrm{GeV}$. The initial conditions of the air shower are the same as in \autoref{fig:polarized_muon_impact_spectra}.} 
  \label{fig:muon_mult_scat_impact}
\end{figure}

In \autoref{fig:muon_mult_scat_impact}, we show the impact of the muon multiple scattering on the sea-level muon angular distributions in a proton-induced air shower, given the same initial conditions as in \cref{fig:polarized_muon_impact_spectra,fig:polarized_muon_impact_angdists}. For a representative example, we focus on the muons with $E_{\mu} \simeq 2\,\mathrm{GeV}$ (i.e., the parents of $\mathcal{O}$(GeV) electron and muon neutrinos). We find that the cumulative effect of muon multiple scattering is a $\sim$$1^{\circ}$ shift of the angular density peak, compared to the air shower evolved without muon multiple scattering. We confirmed that the ``tilt'' seen in the lower panel of \autoref{fig:muon_mult_scat_impact} grows with the distance traversed by the muons. The effect on the neutrino angular distributions was, however, found to be negligible, introducing at most $\mathcal{O}(1\%)$ bias at the sea level if muon multiple scattering was not included in the cascade equations.

\clearpage

\section{Comparison of the configuration settings between CORSIKA and MCEq\label{appendix:cm_benchmarking_details}}

The geomagnetic field and the respective curving of the charged particle trajectories are not implemented in 2D \textsc{MCEq} at the time of writing. We therefore effectively disable the geomagnetic field in \textsc{corsika} by setting $B_x = B_z = 10^{-5}\,\upmu\mathrm{T}$ (\textsc{corsika} requires $|\mathbf{B}| > 0$). In addition, since the 2D \textsc{MCEq} code currently excludes electromagnetic cascades, we also disable the electromagnetic interactions in \textsc{corsika} by setting all electromagnetic flags (``\textsc{elmflg}'') to false.
\color{black}
While the choice of the hadronic models is matched between \textsc{corsika} in \textsc{MCEq}, the switch between the low- and the high-energy regimes means a sharp transition between the two hadronic interaction models in \textsc{corsika} and a smooth interpolation between the models in 2D \textsc{MCEq} (see \cref{appendix:interpolation} for details). Thus, when both energy regimes are covered in a simulated case, small discrepancies between the two codes are possible due to the different implementations of the low-energy/high-energy model transition. When comparing the simulation outputs at the energies below the transition threshold, one further has to be mindful of the different low-energy model versions. We expect the differences due to the low-/high-energy transition implementation and the internal model versions to be smaller than due to a full change of the low-energy interaction model to a different one, which is investigated in \cref{appendix:impact_of_le_hadronic_models}. 

In \textsc{corsika}, the azimuthal angle of the primary particle incidence is fixed at $\varphi_0 = 0$. The height of the first possible interaction of the proton with atmospheric nuclei is set to \SI{112.8}{km} in both \textsc{MCEq} and \textsc{corsika}. The atmospheric density as a function of the slant depth $X$ is modelled according to the Linsley parametrization of the US Standard atmosphere \cite{Heck:1998vt}. In \textsc{MCEq}, the average stopping power of the charged particles due to ionization, bremsstrahlung and pair production is taken from tables provided by the Particle Data Group \cite{Workman:2022ynf, Fedynitch:2018cbl}, whereas the energy derivative $\frac{\partial}{\partial E}$ is approximated as a five-point stencil. In \textsc{corsika}, the average stopping power is calculated analytically via the Bethe-Bloch prescription \cite{Groom:2000xx, Segre:1953xx}, and is directly used to reduce the energy of the charged particles between two propagation steps. The Gauss approximation is employed in both codes for muon angular deflections due to multiple scattering. 

\clearpage

\section{\textsc{MCEq}-\textsc{CORSIKA} benchmarking for a power law cosmic ray spectrum}\label{appendix:power_law_spectrum}

\begin{figure*}[htpb]
  \includegraphics[width=\textwidth]{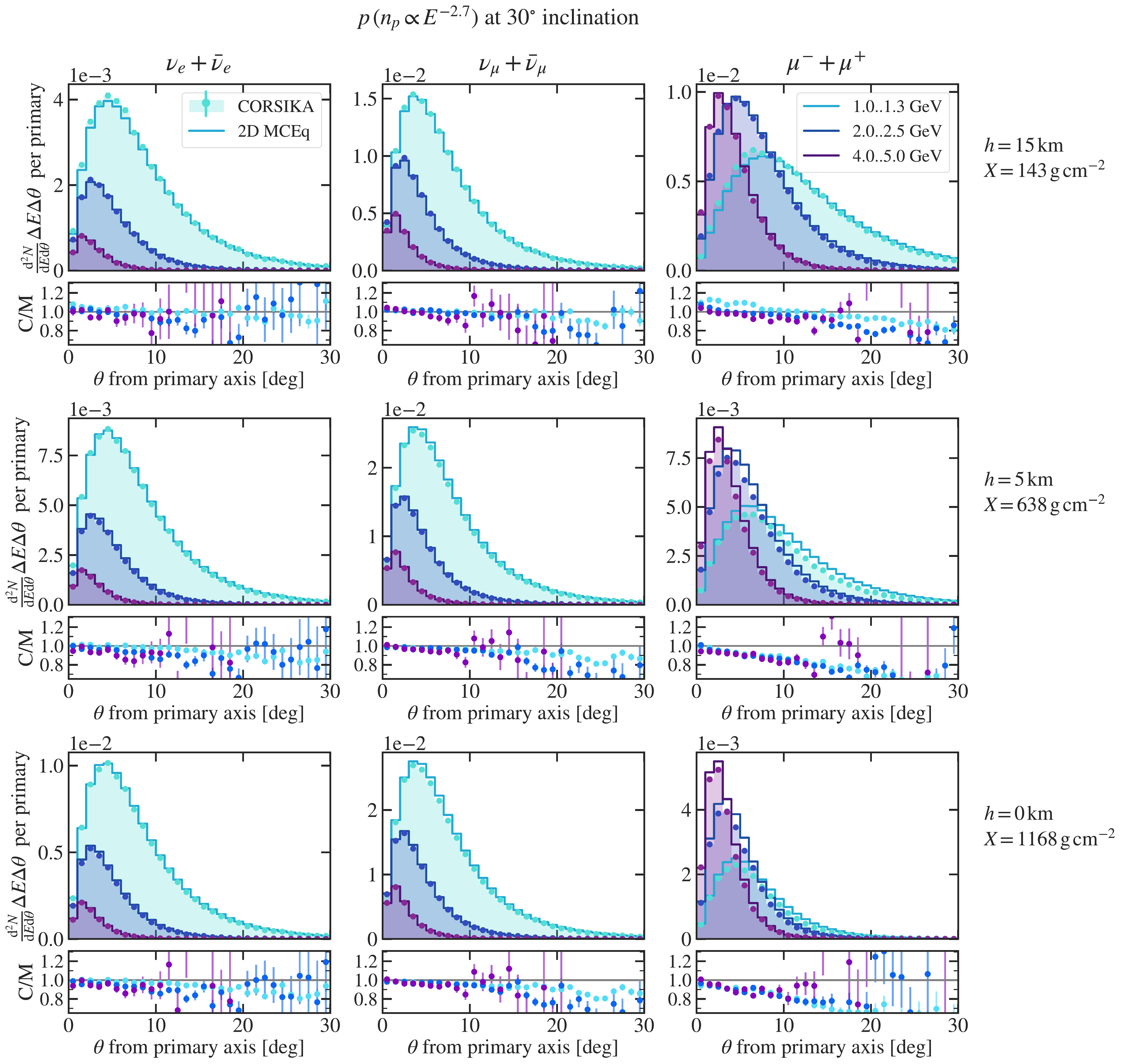}
  \caption{Angular distributions of atmospheric leptons in proton-induced air showers with a power law starting spectrum (see title).}
  \label{fig:ang_dists_spectrum_30deg}
\end{figure*}

\begin{figure*}[htpb]
  \includegraphics[width=\textwidth]{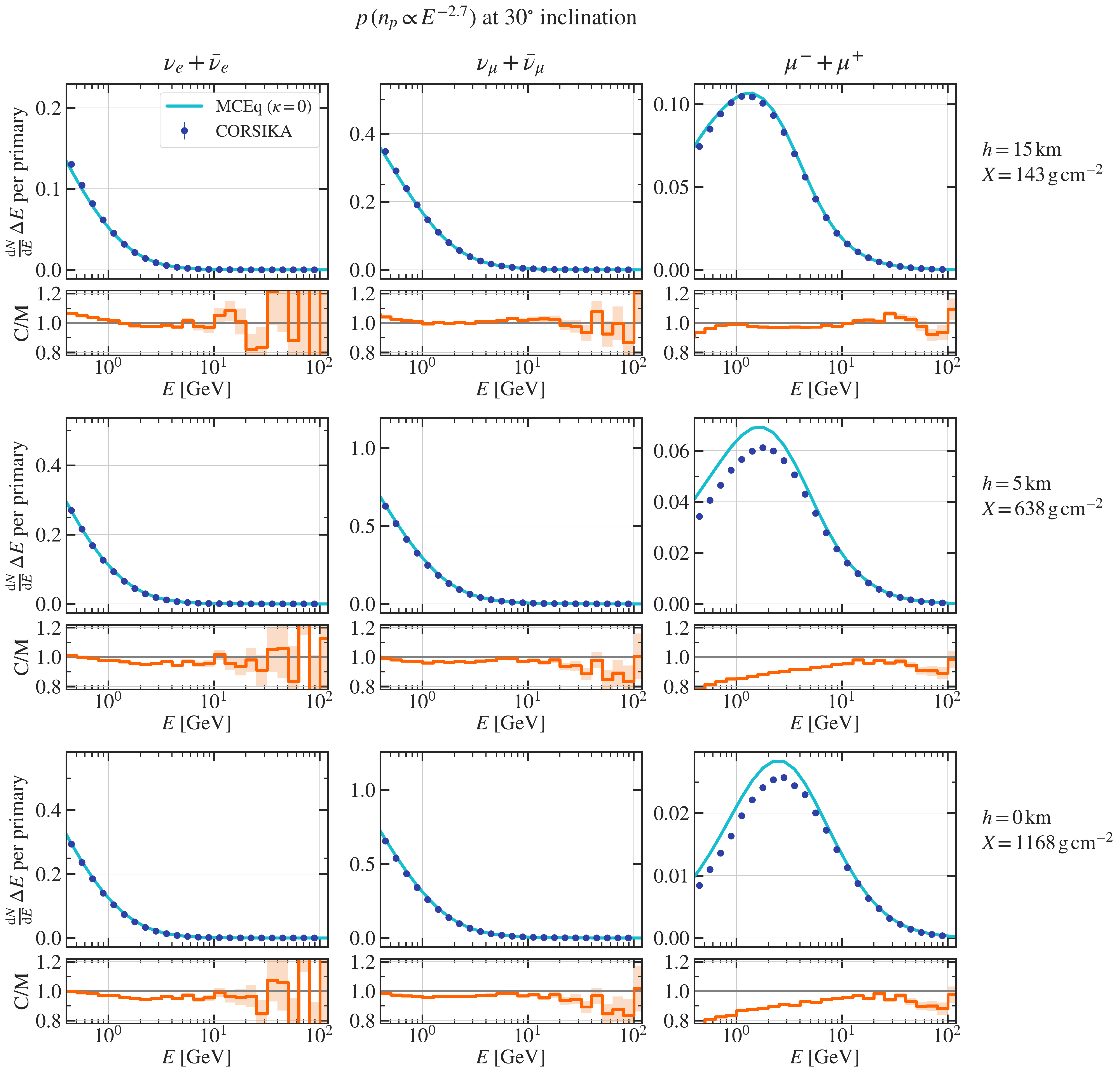}
  \caption{Energy spectra of atmospheric leptons in proton-induced air showers with a power law starting spectrum (see title).}
  \label{fig:en_spectra_spectrum_30deg}
\end{figure*}

\clearpage

\section{\textsc{MCEq}-\textsc{CORSIKA} benchmarking for near-horizontal showers}\label{appendix:horizontal_showers}

\begin{figure*}[htpb]
  \includegraphics[width=\textwidth]{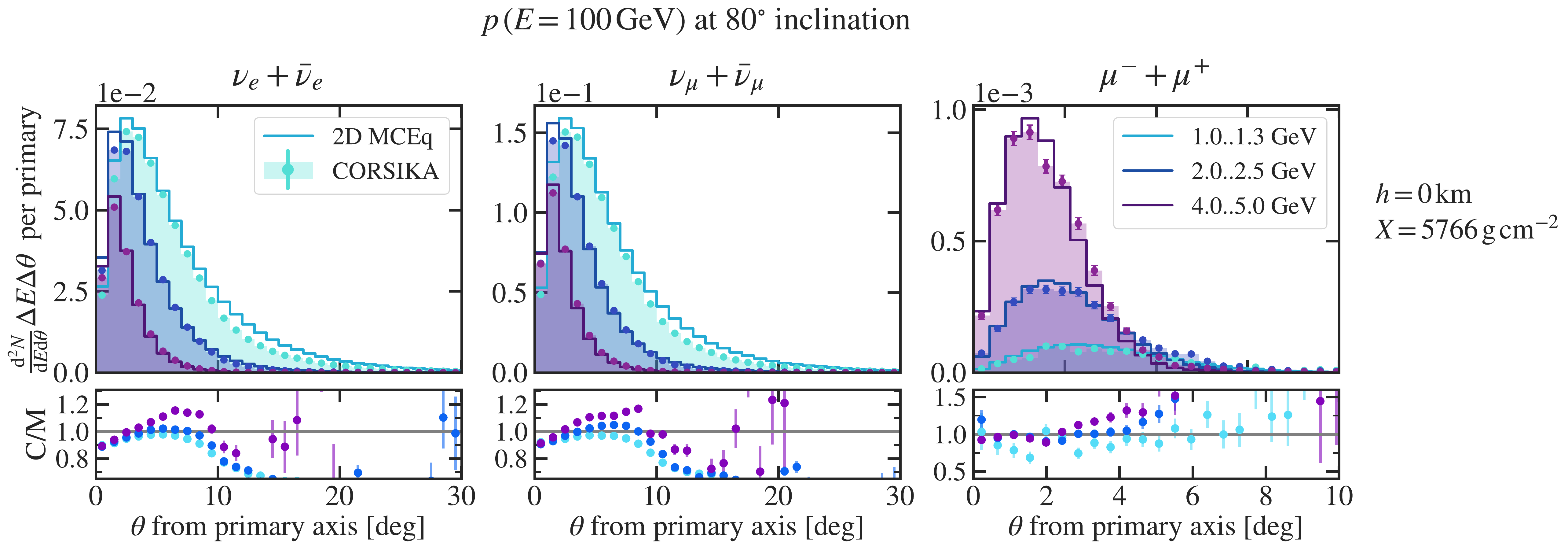}
  \caption{Angular distributions of atmospheric leptons in a proton-induced air shower ($E_0$ = \SI{100}{GeV}, $\theta_0 = 80^{\circ}$), as computed numerically in 2D \textsc{MCEq} (solid lines) and simulated in the \textsc{corsika} Monte Carlo (filled histograms with errorbars).}
  \label{fig:ang_dists_100GeV_80deg}
\end{figure*}

\begin{figure*}[htpb]
  \includegraphics[width=\textwidth]{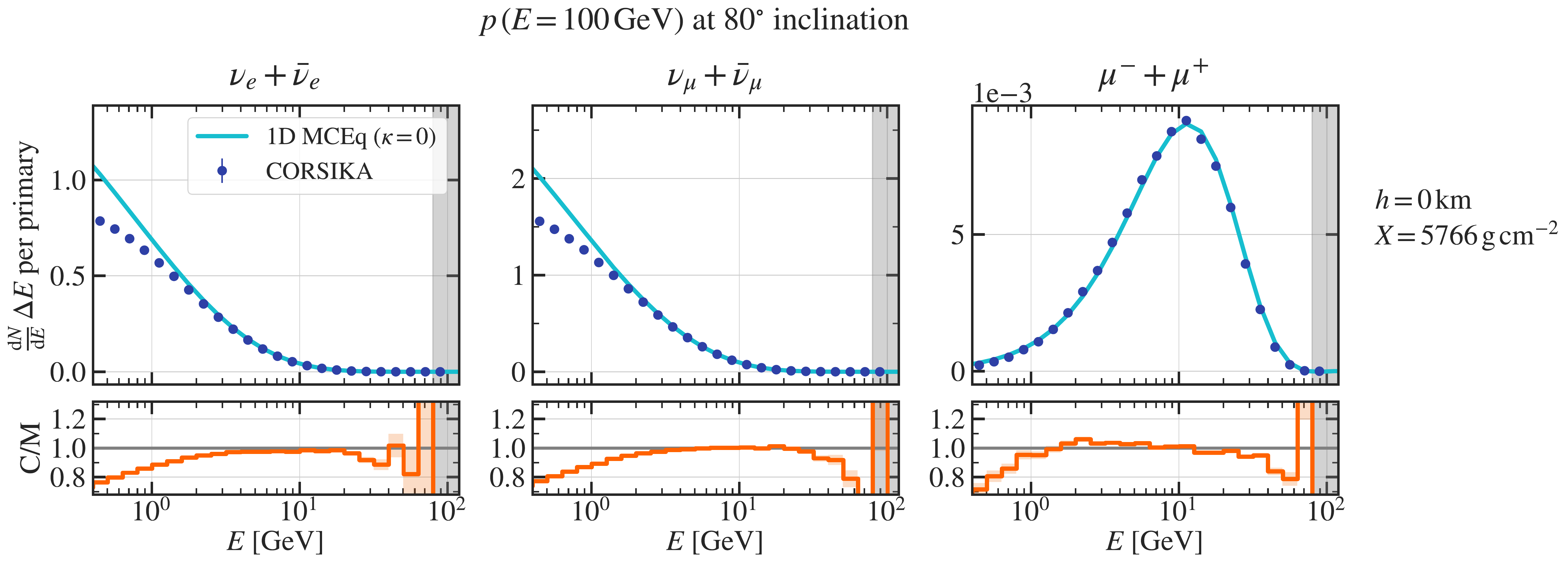}
  \caption{Energy spectra of atmospheric leptons in a proton-induced air shower ($E_{0}$ = \SI{100}{GeV}, $\theta_0 = 80^{\circ}$), as computed numerically in 1D \textsc{MCEq} (solid lines) and simulated in the \textsc{corsika} Monte Carlo (filled markers).}
  \label{fig:en_spectra_100GeV_80deg}
\end{figure*}

\clearpage

\section{Impact of the low-energy hadronic model choice}\label{appendix:impact_of_le_hadronic_models}

\begin{figure*}[htpb]
  \includegraphics[width=\textwidth]{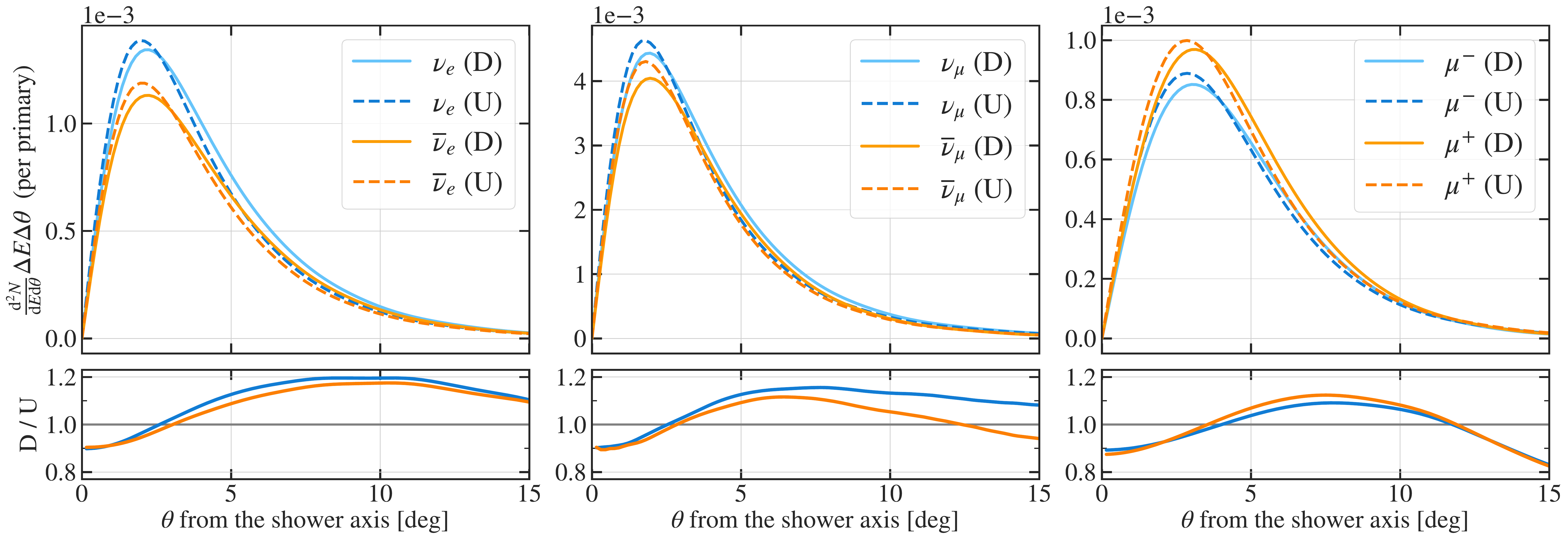}
  \caption{Angular distributions of low-energy atmospheric leptons ($E \simeq 2\,\mathrm{GeV}$) in a proton-induced air shower, as computed numerically in 2D \textsc{MCEq} using two different low-energy hadronic interaction models (\textsc{DPMJet}-III 19.1 (``D'') and \textsc{UrQMD} 3.4 (``U'')). The bottom sub-panel in each plot shows the \textsc{DPMJet}/\textsc{UrQMD} solution ratio.}
  \label{fig:ang_dists_le_model_comparison}
\end{figure*}

\begin{figure*}[htpb]
  \includegraphics[width=\textwidth]{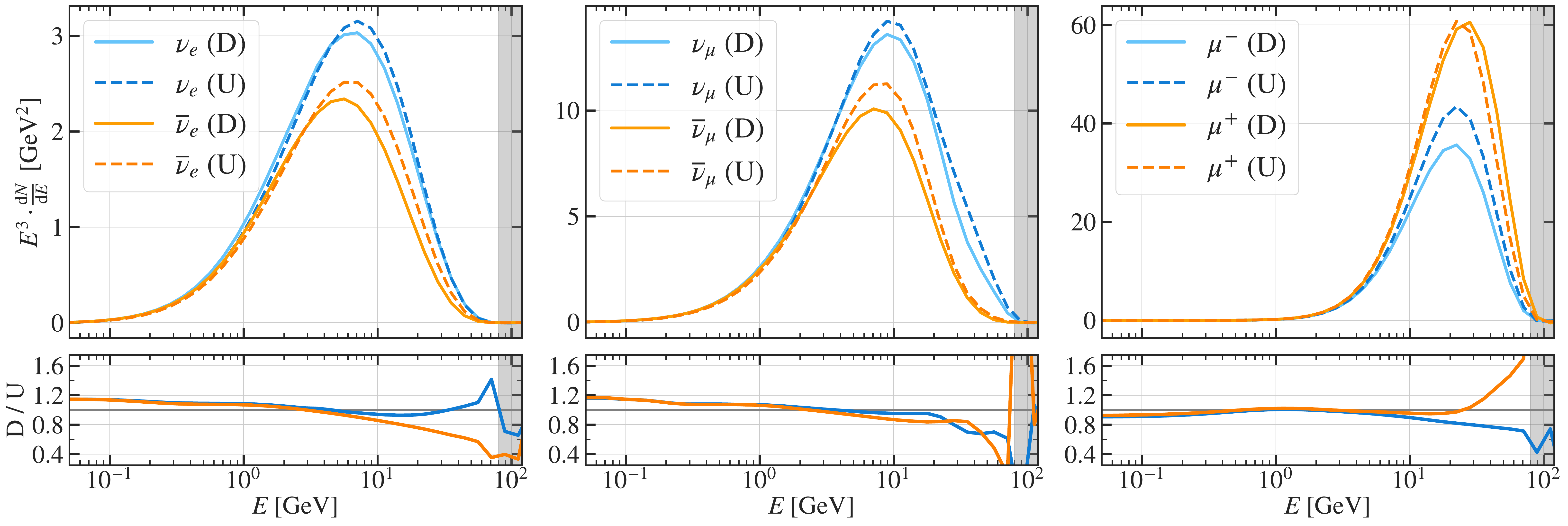}
  \caption{Energy spectra of atmospheric leptons in a proton-induced air shower ($E_{0}$ = \SI{100}{GeV}, $\theta_0 = 30^{\circ}$), as computed numerically in 2D \textsc{MCEq} using two different low-energy hadronic interaction models. The legend follows that of \autoref{fig:ang_dists_le_model_comparison}. The shaded gray band represents the region where the \textsc{MCEq} solution is numerically unstable (see the caption of \autoref{fig:polarized_muon_impact_spectra} for details).}
  \label{fig:en_spectra_le_model_comparison}
\end{figure*}

\newpage
\twocolumngrid

\bibliographystyle{apsrev.bst}

\end{document}